\documentclass[floatfix,prl,twocolumn,aps,superscriptaddress,showpacs]{revtex4-2}

\usepackage[colorlinks=true,linktoc=page,linkcolor=PineGreen,citecolor=magenta,urlcolor=violet]{hyperref}
\usepackage{multirow}
\usepackage{adjustbox}
\usepackage{graphicx}
\usepackage{amsmath}
\usepackage{physics}
\usepackage{amsfonts}
\usepackage{amssymb}
\usepackage{bm}
\usepackage[usenames,dvipsnames]{color}
\usepackage{float}
\usepackage{natbib}
\usepackage{comment}
\hypersetup{
colorlinks=true,
citecolor=blue,
linkcolor=blue,
urlcolor=blue}

\newcommand{\be}{\begin{equation}}
\newcommand{\ee}{\end{equation}}
\newcommand{\bea}{\begin{eqnarray}}
\newcommand{\eea}{\end{eqnarray}}
\newcommand{\bec}{\begin{center}}
\newcommand{\eec}{\end{center}}

\usepackage[dvipsnames]{xcolor}
\usepackage{multirow}
\usepackage{xfrac}
\usepackage{textcomp}
\usepackage{enumerate}
\usepackage{subfigure}
\usepackage{stackengine}
\usepackage{booktabs}
\usepackage{physics}
\usepackage[colorlinks=true,linktoc=page,linkcolor=PineGreen,citecolor=magenta,urlcolor=violet]{hyperref}
\usepackage{newfloat}
\DeclareFloatingEnvironment[name={TABLE},fileext=lsst,listname={List of
	Supplementary Short Tables},within=section, placement=htbp!]{supptable}

\mathchardef\mhyphen="2D 

\newcommand\beq{\begin{equation}}  
\newcommand\eeq{\end{equation}}

\usepackage[normalem]{ulem}
\definecolor{lime}{HTML}{A6CE39}
\usepackage{sidecap,tikz}
\DeclareRobustCommand{\orcidicon}{\hspace{-1.0mm}
	\begin{tikzpicture}
	\draw[lime, fill=lime] (0.0,0.0) 
	circle [radius=0.15] 
	node[white] {{\fontfamily{qag}\selectfont \tiny \,ID}};
	\draw[white, fill=white] (-0.0525,0.095) 
	circle [radius=0.007];
	\end{tikzpicture}
	\hspace{-3.0mm}
}
\foreach \x in {A, ..., Z}{\expandafter\xdef\csname orcid\x\endcsname{\noexpand\href{https://orcid.org/\csname orcidauthor\x\endcsname}
		{\noexpand\orcidicon}}
}

\begin{document}

\title{Paradoxical Topological Soliton Lattice in Anisotropic Frustrated Chiral Magnets}
\author{Sayan Banik\orcidA{}}
\affiliation{School of Physical Sciences, National Institute of Science Education and Research, An OCC of Homi Bhabha National Institute, Jatni-752050, India}
\author{Nikolai S. Kiselev\orcidB{}}
\altaffiliation{email: n.kiselev@fz-juelich.de, aknandy@niser.ac.in}
\affiliation{Peter Gr\"unberg Institute, Forschungszentrum J\"ulich, 52425 J\"{u}lich, Germany}
\author{Ashis K.\ Nandy\orcidC{}}
\altaffiliation{email: n.kiselev@fz-juelich.de, aknandy@niser.ac.in}
\affiliation{School of Physical Sciences, National Institute of Science Education and Research, An OCC of Homi Bhabha National Institute, Jatni-752050, India}


\begin{abstract}
\end{abstract}

\maketitle

\noindent \textbf{Abstract}\\
\noindent Two-dimensional chiral magnets are known to host a variety of skyrmions, characterized by an integer topological charge ($Q \in \mathbb{Z}$). 
However, these systems typically favor uniform lattices as a thermodynamically stable phase composed of either skyrmions ($Q = -1$) or antiskyrmions ($Q = 1$).
In isotropic chiral magnets, skyrmion-antiskyrmion coexistence is typically transient due to mutual annihilation, making the observation of a stable, long-range ordered lattice a significant challenge.  
Here, we address this challenge by demonstrating a skyrmion-antiskyrmion lattice as a magnetic field-induced topological ground state in chiral magnets with competing anisotropic interactions, specifically Dzyaloshinskii-Moriya and frustrated exchange interactions.
This unique lattice exhibits a \textit{net-zero} global topological charge due to the balanced populations of skyrmions and antiskyrmions.
Furthermore, density functional theory and spin-lattice simulations identify 2Fe/InSb(110) as an ideal candidate material for realizing this phase. 
This finding reveals new possibilities for manipulating magnetic solitons and establishes \textit{anisotropic frustrated chiral magnets} as a promising material class for future spintronic applications. 

\vspace{0.4cm}
\noindent \textbf{Introduction}\\
\noindent Topological magnetic solitons~\cite{Kovalev_90} are magnetization field configurations that maintain stable shapes and sizes over time and cannot be continuously transformed into trivial configurations, such as a saturated ferromagnet (FM).
They can also move and interact with each other like ordinary particles. 
Chiral magnets are the most prominent example of a magnetic system exhibiting a wide diversity of topological magnetic solitons~\cite{Rybakov_19, Foster_19, Kuchkin_20i, Kuchkin_20ii, Kuchkin_21, Kuchkin_22homotopy, Kuchkin_23, Zheng_18, Zheng_21, Yang_24embedded, Nandy_NanoLett2020, Heo_SciRep2016, Pham_Science2024}. 
We commonly refer to these solitons as chiral magnetic skyrmions.
In these systems, magnetic skyrmions are stabilized by the competition between Heisenberg exchange interaction and chiral Dzyaloshinskii-Moriya~\cite{Dzyaloshinskii,Moriya} interaction (DMI). 
The existence of statically stable chiral magnetic skyrmions was first predicted theoretically~\cite{Bogdanov_89} and later confirmed experimentally through both direct~\cite{Yu_10, Yu_11, Heinze_NatPhys2011, Seki_12multiferroic, Romming_13, Kezsmarki_15polar, Nayak-Nature2017, Grenz_17} and indirect~\cite{Muehlbauer-Science2009} observations of their hexagonal lattice arrangement in various compounds.

The diversity of magnetic skyrmions can be classified using the homotopy group concept.
In the case of two-dimension (2D), the classifying group is the second homotopy group with respect to the $\mathbb{S}^2$ sphere, which is known to be isomorphic to the group of integers, $\pi_2(\mathbb{S}^2) = \mathbb{Z}$. 
The latter means that each topological soliton can be associated with a specific integer called the skyrmion topological charge:
\begin{equation}
Q = \frac{1}{4\pi}\int_{\Omega} \mathbf{m} \cdot \left[\partial_x \mathbf{m} \times \partial_y \mathbf{m}\right] \, \mathrm{d}x \, \mathrm{d}y, 
\label{eq_Q}
\end{equation}
where $\mathbf{m}(x,y) = \mathbf{M} / |\mathbf{M}|$ is the unit vector field of magnetization, and $\Omega$ is the skyrmion localization area chosen such that at its boundary $\partial \Omega$, the magnetization field points in one direction, $\mathbf{m}(\partial \Omega) = \mathbf{m}_0$.
Following the standard convention for sign definiteness in \eqref{eq_Q}, we assume a right-handed Cartesian coordinate system and that magnetization at the boundary satisfies the criterion, $\mathbf{m}_0 \cdot \mathbf{e}_z > 0$.
For details, see Refs.~\cite{Rybakov_19, Zheng_22}.
%
\begin{figure*}
    \centering
    \includegraphics[width=1\linewidth]{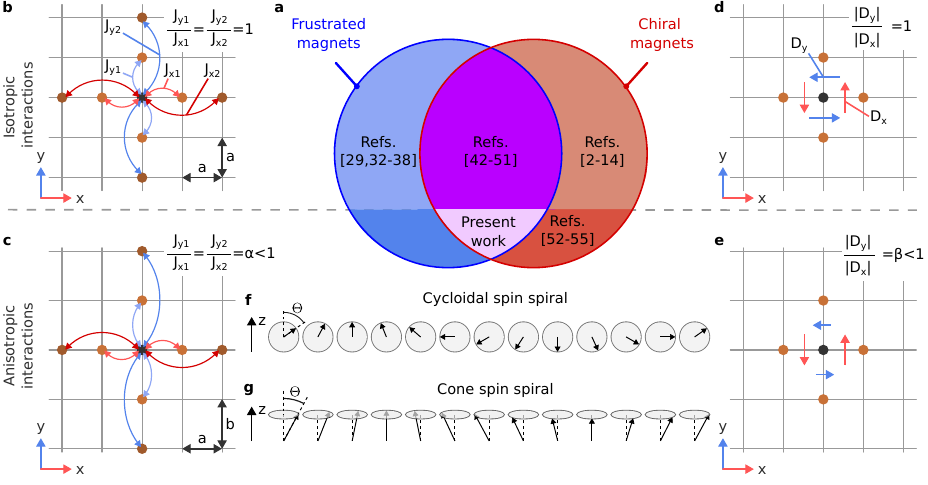}
\caption{\textbf{Diversity of systems with frustrated exchange and chiral interactions.}
\textbf{a,} The diagram illustrates the diversity of systems with frustrated Heisenberg exchange interactions and DMI, including both isotropic and anisotropic cases. The case studied in the present work corresponds to the system with anisotropic interaction parameters, $i.e.$, the \textit{anisotropic frustrated chiral magnet}.
\textbf{b,} The schematic representation of the minimal model for the 2D magnet with isotropic frustrated exchange interactions on a square lattice where nearest and the next after-nearest neighbor exchange constants have opposite sign, $J_1>0$ (ferromagnetic) and $J_2<0$ (antiferromagnetic), see the spin-lattice Hamiltonian~\eqref{spin_hamiltonian} in Methods.
\textbf{c,} The schematic representation of a rectangular lattice with anisotropic frustrated exchange interactions, characterized by unequal exchange coupling constants along the $x$- and $y$-directions. 
\textbf{d,} The DMI vectors between the nearest neighbors on a square lattice for the interfacial type DMI. For an isotropic system, the absolute values of the DMI vectors in both $x$ and $y$ directions are identical. 
\textbf{e,} The system with anisotropic DMI on a rectangular lattice, where the magnitude of the DMI differs along the $x$- and $y$-directions.
\textbf{f,} A cycloidal spin spiral, often resulting from interfacial DMI in chiral magnets, characterized by a varying polar angle $\Theta$ with respect to the $z$-axis along the propagation direction. 
\textbf{g,} A cone spin spiral, characterized by a fixed polar angle $\Theta$ and a varying azimuthal angle along the propagation direction.
}
\label{fig1}
\end{figure*}

The solitons of opposite topological charges are typically called skyrmions ($Q=-1$) and antiskyrmions ($Q=1$).
In isotropic systems, depending on the crystallographic symmetry, only one type of particle, either skyrmion or antiskyrmion, is energetically most favorable.
These particles form periodic lattices with long-range order in a specific range of the external magnetic field.
Such lattices, composed of one sort of particle, either skyrmions or antiskyrmions, have been experimentally observed in various compounds~\cite{Muehlbauer-Science2009, Yu_10, Yu_11, Heinze_NatPhys2011, Seki_12multiferroic, Romming_13, Kezsmarki_15polar, Nayak-Nature2017, Grenz_17, ASk_d2d}.

In this work, we report the paradoxical phenomenon of a stable regular lattice composed of both skyrmions and antiskyrmions. 
The paradox stems from the fact that, under normal conditions, skyrmion and antiskyrmion act as particle-antiparticle pairs, exhibiting mutual annihilation or spontaneous generation phenomena~\cite{Dupe_18trochoidal, Zheng_22}, much like an electron and a positron. 
We demonstrate that in systems with anisotropic interactions, including DMI and frustrated exchange, the skyrmion-antiskyrmion lattice (S-AL) carrying \textit{net-zero} topological charge has an equilibrium period and becomes the lowest energy state within a specific range of external magnetic fields.
Our claims are supported by an anisotropic micromagnetic model, followed by density functional theory (DFT) calculations for a realistic system and corresponding spin lattice model analyses. 
In particular, we demonstrate a prototype 2D chiral magnet in which the discussed phenomenon occurs.
This discovery represents a vital element of the general theory of topological solitons in chiral magnetic systems and beyond.

The paper is organized as follows. First, we introduce the minimal micromagnetic model that predicts the S-AL phase as the ground state in 2D magnetic systems{--}a class of magnets we term \textit{anisotropic frustrated chiral magnets}. 
We discuss the connection between our model and other systems previously studied in the literature, highlighting the role of specific model parameters in stabilizing the S-AL. 
Additionally, we report the formation of stable clusters composed of equal and unequal numbers of skyrmions and antiskyrmions, demonstrating that annihilation is not a necessary outcome.
Following this, we present a realistic heterostructure{--}a Fe double layer on an InSb(110) semiconductor substrate{--}where our DFT calculations and analysis of the corresponding spin lattice model predict the S-AL phase as the energetically favorable lowest energy state in a wide range of applied magnetic fields.

\vspace{0.4cm}
\noindent \textbf{2D magnet with anisotropic interactions}\\
\noindent We consider the model of 2D magnets, which is composed of Zeeman energy term, frustrated exchange interaction, and DMI. 
The micromagnetic energy functional can be written as follows:
\begin{equation}
E= \int \left\{\mathcal{E}_\mathrm{z}(\mathbf{m})+\mathcal{E}_\mathrm{e}(\mathbf{m})+\mathcal{E}_\mathrm{d}(\mathbf{m}) \right\} t\, \mathrm{d}x \, \mathrm{d}y,
\label{Micro_Ham}
\end{equation}
where magnetization field $\mathbf{m}\equiv\mathbf{m}(x,y)$ is assumed to be homogeneous along the slab thickness $t$.
In the Zeeman energy term,
$\mathcal{E}_\mathrm{z}=-|M|\mathbf{m}\cdot \mathbf{B}_\mathrm{ext}$, the external field is always assumed to be perpendicular to the slab.
$\mathcal{E}_\mathrm{e}$ and $\mathcal{E}_\mathrm{d}$ stand for the energy density of Heisenberg exchange interaction and DMI, respectively.

It is important to note that considering only anisotropic DMI is often insufficient for accurately describing real systems. 
A comprehensive and consistent model should account for both exchange and DMI being anisotropic, reflecting the system's inherent symmetry breaking, as shown in Figs.~\ref{fig1}\textbf{c} and \textbf{e}.
Because of that, in our model, the Heisenberg exchange interaction energy with second-order and fourth-order terms is represented by 
\begin{equation}
\mathcal{E}_\mathrm{e}\!=\!\mathcal{A}_x\!\Biggl(\!\frac{\partial \mathbf{m}}{\partial x}\!\Biggr)^{\!\!2} 
\!+\mathcal{A}_y\!\Biggl(\!\frac{\partial \mathbf{m}}{\partial y}\!\Biggr)^{\!\!2} 
\!+\mathcal{B}_x\!\left(\!\frac{\partial^2\mathbf{m}}{\partial x^2}\!\right)^{\!\!2}
\!+\mathcal{B}_y\!\left(\!\frac{\partial^2\mathbf{m}}{\partial y^2}\!\right)^{\!\!2}
\label{Micro_Ham_Ex}
\end{equation}
This energy term represents one of the limiting cases of a more general model,~\cite{Rybakov_2022} describing frustrated magnets. 
In terms of the atomistic spin-lattice model, the exchange energy term can be approximated by the Heisenberg exchange between the nearest and the next after-nearest neighbors, as depicted in Figs.~\ref{fig1}\textbf{b} and \textbf{c}.
The corresponding micromagnetic terms can be obtained by Taylor expansion of the spin-lattice Hamiltonian with respect to the lattice parameter~\cite{Rybakov_2022}. 
The micromagnetic terms and spin-lattice coupling constants in a square lattice are related as follows: $\mathcal{A}=(J_1/2+2J_2)/a$ and $\mathcal{B}=-(J_1/96+J_2/6)a$, where $a$ is the lattice constant. More details can be found in \textbf{Supplementary Note 1}~\cite{supplementary}.

In the isotropic case, where $\mathcal{A}_x=\mathcal{A}_y=\mathcal{A} > 0$ and $\mathcal{B}_x=\mathcal{B}_y=\mathcal{B} > 0$, the minimum energy of the term \eqref{Micro_Ham_Ex} corresponds to a collinear FM state.
In the case of frustrated exchange interaction, when $\mathcal{A}<0$ and $\mathcal{B}>0$, the ground state of the system is a spin spiral (SS) that is degenerate with respect to the rotation of magnetization $\mathbf{m}$ about any arbitrary axis.
The equilibrium period of such a flat SS is defined by the ratio between exchange coupling constants, $L_\mathrm{H} = 4\pi\sqrt{{\mathcal{B}}/{|\mathcal{A}|}}$.
Under the external magnetic field, its degeneracy with respect to an arbitrary rotational axis is broken.
In a range of external magnetic fields, $0\leq B_\mathrm{ext}<B_\mathrm{c}$, the conical SS (cone-SS) is the lowest energy state.
So, there exists a critical field, $B_\mathrm{c} = \mathcal{A}^2/(4M_\mathrm{s}\mathcal{B})$, above which the cone-SS phase continuously converges to a saturated FM state.
Notably, the period $L_\mathrm{H}$ of the cone-SS does not depend on the strength of the applied magnetic field.
The exchange term \eqref{Micro_Ham_Ex} can also be seen as a limiting case of the interaction studied in Refs.~\cite{Ivanov1986, Ivanov1990}. 
In these seminal works, the stability of magnetic skyrmions, driven by higher-order exchange interactions{--} which is now often referred to as exchange frustration{--}was first predicted.
Over the years, the concept of skyrmions stabilized by exchange frustration has evolved gradually~\cite{Abanov1998, Kirakosyan2006, Leonov_15, Zhang_2017, Dupe_18trochoidal, speight2020skyrmions}.
However, this type of skyrmions has not garnered as much attention as DMI-stabilized skyrmions in pure chiral magnets.

In the model \eqref{Micro_Ham}, we consider the interfacial type DMI, which, in the micromagnetic limit, can be written as
\begin{equation}
	\mathcal{E}_\mathrm{d}= \mathcal{D}_x\Lambda^{(x)}_{xz}+\mathcal{D}_y\Lambda^{(y)}_{yz},
	\label{Micro_Ham_DMI}
\end{equation}
where Lifshitz invariants are defined as follows,
\begin{equation}
	\Lambda_{ij}^{(k)} = m_i\frac{\partial m_j}{\partial k} - m_j\frac{\partial m_i}{\partial k}.  
	\label{LI}
\end{equation}
In the atomistic spin-lattice model, it corresponds to the case where the DMI vector, $\mathbf{D}$, is perpendicular to the $\mathbf{r}$-vector between the interacting spins, as depicted in Figs.~\ref{fig1}\textbf{d} and \textbf{e}.
Such orientations of the DMI vectors are common in all 2D systems with $C_{nv}$ symmetry.
The micromagnetic DMI constants in the case of a square lattice can be expressed in terms of the spin-lattice model parameters as follows: $\mathcal{D}_x=aD_x$ and $\mathcal{D}_y=aD_y$.
The full range of systems described by model \eqref{Micro_Ham} is illustrated in the diagram shown in Fig.~\ref{fig1}\textbf{a}. 
Here, this diagram visually distinguishes between three distinct magnetic regimes: pure frustrated magnets (blue), pure chiral magnets (red), and systems with competing exchange frustration and DMI (magenta).

Our model, incorporating interfacial-type DMI and exchange frustration, supports two distinct types of SSs. 
When DMI dominates, the system stabilizes into a cycloidal-SS, as shown in Fig.~\ref{fig1}\textbf{f}. In the presence of a nonzero magnetic field and dominating exchange frustration, the cone-SS depicted in Fig.~\ref{fig1}\textbf{g}  emerges as the lowest energy state. 
These SSs can be characterized by two fundamental properties: chirality, defined as \( \sim \mathbf{q} \cdot (\nabla \times \mathbf{m}) \), and polarity, given by $ \sim| \mathbf{q} \times (\nabla \times \mathbf{m}) |$, where \( \mathbf{q} \) denotes the SS wave vector. The cycloidal-SS in Fig.~\ref{fig1}\textbf{f} exhibits zero chirality but nonzero polarity. In contrast, the cone-SS in Fig.~\ref{fig1}\textbf{g} shows both nonzero chirality and polarity.  
For reference, helical SSs, characteristic of systems with bulk-type DMI and not shown here, display nonzero chirality but zero polarity.

\begin{figure*}
    \centering
    \includegraphics[width=1.0\linewidth]{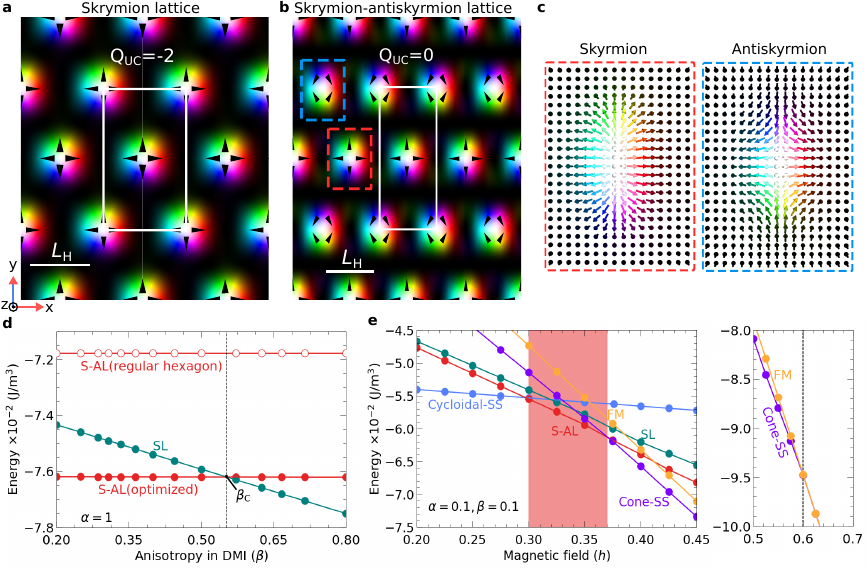}
\caption{\textbf{Achieving S-AL ground states within model~\eqref{Micro_Ham} using anisotropic parameters:} Through micromagnetic simulations utilizing direct energy minimization of our model, we obtain optimized unit cells (white rectangular boxes) for both SL and S-AL phases, as presented in \textbf{a} and \textbf{b}, respectively. 
The optimization scheme is detailed in fig.~S9~\cite{supplementary}. 
The magnetization vector field is visualized by standard color code.
Here, $Q_\textrm{UC}=0$ signifies the rectangular unit cell which represents the \textit{net-zero} topological charge lattice.
\textbf{c,} Left and right spin textures depict individual skyrmion and antiskyrmion, respectively, obtained from the optimized S-AL configuration in \textbf{b}. 
\textbf{d,} Energy profiles for the SL and S-AL phases as a function of $\beta$. The S-AL phase is energetically favored over the SL phase for $\beta < \beta_\textrm{c}~(\approx 0.55$). 
The energy curves of the SL and S-AL phases intersect at the critical value $\beta_\textrm{c}$. 
The significant energy gain observed in the S-AL phase through shape optimization ($\theta$ optimization) is the primary reason for the high value of $\beta_\textrm{c}$. 
All calculations are performed under a constant external magnetic field of $h=0.35$.
\textbf{e,} Energies for all competing non-collinear states{--}cycloidal-SS, cone-SS, SL, and S-AL{--}as well as saturated FM state are plotted as a function of $h$ for \textit{anisotropic frustrated chiral magnets}. Here, we fix both $\alpha$ and $\beta$ to 0.1. 
The S-AL phase is identified as the ground state within the red-shaded regions, corresponding to the lowest energy states for specific ranges of $h$.
For a comparison of the system's behavior at higher anisotropy, $\beta > \beta_\textrm{c}$, see fig.~S10~\cite{supplementary}.}
    \label{fig:MicroMag_intA}
\end{figure*}

The key parameters of our model define the anisotropy in the exchange interaction and DMI.
The parameter $\alpha={\mathcal{A}_y}/{\mathcal{A}_x}={\mathcal{B}_y}/{\mathcal{B}_x}$ defines the anisotropy in exchange interaction.
In a more general approach, one could consider the case when $\dfrac{\mathcal{A}_y}{\mathcal{A}_x}\neq\dfrac{\mathcal{B}_y}{\mathcal{B}_x}$, but here we ignore this option for simplicity of the model.
The anisotropy in DMI is defined by parameter $\beta ={\mathcal{D}_y}/{\mathcal{D}_x}$. 
Note that $\beta$ depends only on the absolute values of the DMI vectors, while their directions are fixed by the lattice symmetry [see Figs.~\ref{fig1}\textbf{d} and \textbf{e}].
In the following, we show that anisotropic DMI, $\beta<1$, is the key ingredient for stabilization of the S-AL depicted in Fig.~\ref{fig:MicroMag_intA}\textbf{b}.

The parameters $\alpha$ and $\beta$ can be freely adjusted within the range $0 \leq \alpha \leq \infty$ and $0 \leq \beta \leq \infty$.
However, for definiteness, here, we assume that $\alpha \in [0,1]$ and $\beta\in [0,1]$. Therefore, the coupling strengths along the $y$-axis are always weaker than those along the $x$-axis.
Hereafter, the inherent anisotropy, with the $x$-axis as the \textit{strong} axis and the $y$-axis as the \textit{weak} axis, 
gives rise to a distinct class of chiral magnets, specifically designated as \textit{anisotropic frustrated chiral magnets}.
%
The elongated shapes of skyrmions and antiskyrmions, clearly visible in Fig~\ref{fig:MicroMag_intA}\textbf{c}, are a direct consequence of the anisotropic properties of the exchange interaction and DMI in our model.

In the isotropic case with $\alpha=1$, $\beta=1$, and positive exchange coupling constants ($\mathcal{A}>0$ and $\mathcal{B}\geq0$), model \eqref{Micro_Ham} simplifies to the extensively studied model of isotropic chiral magnets~\cite{Bogdanov_89, Bogdanov_99, Bogdanov_1994, Bogdanov_1994JMMM, Rybakov_19, Foster_19, Yang_24embedded, Kuchkin_20i, Kuchkin_20ii, Kuchkin_21, Kuchkin_22homotopy, Kuchkin_23}.
To date, only systems with isotropic frustrated exchange and isotropic DMI ($\alpha=\beta=1$) have been explored in the literature~\cite{Dupe_NatCommun2014, Dupe-NatCommun2016, Rozsa_16, Nandy_PRL2016, yuan2017skyrmions, rozsa2017formation, vonMalottki2019, zhang2020static, goerzen2023lifetime, Nandy-PRBL2024}. 
These systems are represented by the dark magenta region in Fig.~\ref{fig1}\textbf{a}.
Most of the parameter space within the isotropic regime can be attributed to known limiting cases of model~\eqref{Micro_Ham}, which we systematically examine in \textbf{Supplementary Note 2}~\cite{supplementary}.

In contrast, anisotropic systems have been primarily studied in the context of DMI only~\cite{Hoffmann-NatCommun2017, Paterson2019, Du_23mono, kuchkin_23monoaxial}.
This includes the limiting case of strongly anisotropic systems corresponding to so-called \textit{monoaxial} chiral magnets~\cite{kuchkin_23monoaxial, Togawa_12, Du_23mono}.
The models featuring both anisotropic DMI and anisotropic exchange, as considered in this work with terms \eqref{Micro_Ham_Ex} and \eqref{Micro_Ham_DMI}, remain largely unexplored.
Finally, it is important to emphasize that our model does not include magnetocrystalline anisotropy or any other form of spin orientation anisotropy, as they are not essential for the phenomena discussed here. 
However, to ensure a comprehensive analysis, we examine the role of magnetocrystalline anisotropy in stabilizing the S-AL phase, including a 2D material example: the 2Fe/InSb(110) heterostructure.

\vspace{0.4cm}
\noindent \textbf{Model parameters}

\noindent Below, we consider the solutions of the model \eqref{Micro_Ham} with $\mathcal{A}_{(x)y} < 0$ and $\mathcal{B}_{x(y)} > 0$, assuming for definiteness that $\mathcal{D}_{x(y)} > 0$.
To facilitate further analysis, we introduce two parameters, $L_\textrm{D}$ and $L_\textrm{H}$, representing the equilibrium period of the spiral state in two limiting cases. 
When $\mathcal{A} \to 0$, the equilibrium period of chiral spiral state is $L_\textrm{D} = 2\pi \sqrt[3]{16\mathcal{B}_{x}/\mathcal{D}_{x}}$, and when $\mathcal{D} \to 0$, the exchange frustration driven spiral has period $L_\textrm{H} = 4\pi\sqrt{\mathcal{B}_{x}/|\mathcal{A}_{x}|}$ (see \textbf{Supplementary Note 1}~\cite{supplementary}). 
In the first approximation, the ratio between $L_\mathrm{H}$ and $L_\mathrm{D}$ characterizes the relative contributions of the frustrated exchange and DMI terms to the stability of SS and other noncollinear phases.
Frustrated exchange dominates DMI when $L_\mathrm{H}<L_\mathrm{D}$ and vice versa.
The reduced magnetic field, $h=B_\textrm{ext}/B_\textrm{c}$, is provided in units relative to the critical field, $B_\mathrm{c} = \mathcal{A}_x^2/(4M_\mathrm{s}\mathcal{B}_x)$.

Following the standard approach, we perform energy minimization of the functional \eqref{Micro_Ham} for various spin configurations with optimized parameters across different values of the external magnetic field. 
For example, the SS state is optimized with respect to its period and propagation direction. 
In contrast, various lattices of magnetic skyrmions are optimized in terms of the shape and size of their unit cells. 
We then compare the energy densities of all phases to determine the lowest energy state as a function of external magnetic fields. 
The energy minimization was performed using the mumax code~\cite{mumax} (see Methods), and the corresponding script is provided in the Supplementary Data.

\vspace{0.4 cm}
\noindent \textbf{Results and Discussions}

\vspace{0.1 cm}
\noindent \textbf{Micromagnetic model for S-AL in frustrated chiral magnets}

\noindent To begin, we illustrate the results of the energy minimization for the case of isotropic exchange, $\alpha=1$, but anisotropic DMI, $\beta<1$.
In these calculations we keep fixed value of $\mathcal{A}_x = \mathcal{A}_y=-10^{-17}$ J/m, while other parameters ($\mathcal{B}_x$, $\mathcal{B}_y$, $\mathcal{D}_x$, $\mathcal{D}_y$) are defined by parameter $\beta$ and fixed values of $L_\textrm{H}=50$ nm, and $L_\textrm{D}=100$ nm ($L_\mathrm{H}/L_\mathrm{D}=0.5$), see also \textbf{Supplementary Note 1~\cite{supplementary}}.
Figures~\ref{fig:MicroMag_intA}\textbf{a} and \textbf{b} depict the equilibrium SL and S-AL configurations, respectively.  
As detailed in fig.~S9~\cite{supplementary}, the equilibrium lattice phase corresponds to the energy minimum in the parameter space defined by the dimensions of the rectangular unit cell in both directions.
The equilibrium SL and S-AL configurations shown in Figs.~\ref{fig:MicroMag_intA}\textbf{a} and \textbf{b} correspond to a magnetic field of $h=0.35$ applied along the negative $z$-axis.
The S-AL phase exhibits a balanced population of $Q=\pm 1$ topological charges, forming a lattice with \textit{net-zero} topological charge ($Q_\textrm{UC}=0$) per unit cell.
In this phase, skyrmions and antiskyrmions elongate (Fig.~\ref{fig:MicroMag_intA}\textbf{c}), causing noticeable distortion from the ideal hexagonal lattice.
In contrast, the skyrmion elongation in the SL phase at $\beta=0.1$ is minimal, and the resulting lattice distortion is not noticeable. 
Moreover, the equilibrium unit cell of S-AL is slightly larger than that of SL.
As illustrated in Fig.~\ref{fig:MicroMag_intA}\textbf{d}, when both the SL and the S-AL phases are constrained to a regular hexagonal symmetry and optimized solely with respect to the scaling parameter, the energy of the S-AL phase increases significantly.
The scaling parameter refers to the uniform scaling of the rectangular unit cell in both directions. 
This emphasizes the necessity of optimizing both lattice parameters to achieve the lowest-energy lattice shape.  

Most importantly, Fig.~\ref{fig:MicroMag_intA}\textbf{d} shows that with varying anisotropy in the DMI, the energy of S-AL can become lower than that of SL.
In this case, the critical value of the anisotropy parameter at which the energies of SL and S-AL are equal is around $\beta_\mathrm{c}\approx0.55$, while, strictly speaking, $\beta_\mathrm{c}$ is a function of $h$. This is detailed in fig.~S11d~\cite{supplementary}.
We conclude that anisotropic DMI can reverse the energy balance between SL and S-AL.
%
%
Anisotropic DMI alone does not make the S-AL phase ground state.  
In the case of isotropic exchange ($\alpha = 1$), both SL and S-AL remain metastable, meaning their energies are higher than those of other phases.
Thus, while anisotropic DMI is necessary in our model, it is not sufficient to make S-AL the lowest energy state within a specific field range.

\begin{figure*}
    \centering
    \includegraphics[width=1.0\linewidth]{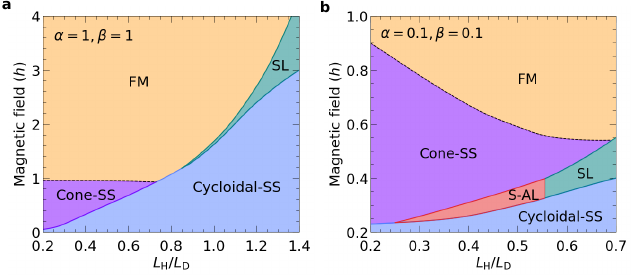}
    \caption{\textbf{The phase diagrams in the $L_\mathrm{H}/L_\mathrm{D}$-$h$ plane.} \textbf{a,} The phase diagram for the isotropic magnets. The parameter space is clearly divided into two regions: one dominated by exchange frustration and the other by DMI. Decreasing the ratio $L_\mathrm{H}/L_\mathrm{D}$ (enhancing exchange frustration) stabilizes the cycloidal-SS phase in the absence of external magnetic field $h$. Further, under an external field $h$, a cone-SS phase appears, with SL and S-AL as metastable states. In this region, the system exhibits three energetically favored phases: cycloidal-SS, cone-SS, and FM. The dashed line between cone-SS and FM denotes the second-order phase transition between them. Dominant DMI interactions are evidenced by the expansion of the cycloidal-SS ground state region at higher $L_\mathrm{H}/L_\mathrm{D}$ ratios. Typically, the cone-SS phase disappears under an external field in the strong DMI limit, while the cycloidal-SS phase undergoes a first-order phase transition to the SL phase. \textbf{b,} The phase diagram for the \textit{anisotropic frustrated chiral magnet}. At $L_\mathrm{H}/L_\mathrm{D}\approx0.56$, the SL and S-AL phases exhibit equal energy across a range of $h$, defining a boundary between these two distinct lattice phases. 
    The SL phase becomes energetically favorable compared to the S-AL phase upon increasing further $L_\mathrm{H}/L_\mathrm{D}$ ratios. This can be attributed to the fact that the critical value of $\beta_c$ remains below 0.1 in this regime.
    }
    \label{fig:MicroMag_intB}
\end{figure*}

In the case of anisotropic interactions, the energy dependencies of the phases differ.
For simplicity, but without loss of generality, we consider the case of $\alpha=\beta$.
In Fig.~\ref{fig:MicroMag_intA}\textbf{e}, we present the energies of the various magnetic phases as a function of applied magnetic field, calculated for the \textit{anisotropic frustrated chiral magnets} with $\alpha=\beta=0.1$. 
%
At low field strengths, the cycloidal-SS phase is the ground state, and it undergoes a first-order phase transition to the S-AL phase at a critical field of $h \approx 0.3$.
A subsequent first-order phase transition occurs at $h \approx 0.37$, beyond which the cone-SS phase emerges as the ground state. Consequently, the S-AL phase is stabilized within a well-defined range of magnetic fields, emphasizing the crucial role of anisotropic interactions in determining the ground state.
Noticeably, the ordinary SL remains a metastable state throughout the entire field range, with its energy curve nearly parallel to that of the S-AL phase. 
As the magnetic field is further increased, a second-order phase transition takes place at approximately $h \approx 0.6$, leading to a transition from the cone-SS to the saturated FM state (as indicated by the vertical line in the rightmost panel of Fig.~\ref{fig:MicroMag_intA}\textbf{e}.)
To gain deeper insights into the influence of anisotropic exchange interactions on the stability of the S-AL phase, we present additional analyses in figs.~S10 and S11~\cite{supplementary}. As in fig.~S12~\cite{supplementary}, spin texture elongation within the S-AL phase is a direct consequence of the system seeking the lowest energy state. Within the constraints of our chosen DMI configuration, elongation along the $y$-direction is energetically most favorable with a deformed hexagonal lattice. 

We conclude that the S-AL phase emerges as the lowest energy state within a specific magnetic field range only when $\beta < \beta_\textrm{c}$, and the anisotropy parameter $\alpha$ is below a critical value.
In the case of Fig.~\ref{fig:MicroMag_intA}\textbf{e}, for $\beta=0.1$, the critical value of anisotropy in exchange interactions $\alpha$ is found to be $\approx 0.26$.
Reducing $\alpha$ below this critical value favors the S-AL phase as the ground state in a broader window of applied fields.

It is worth noting that, in general, $\beta_\textrm{c}$ is a function of the model parameters.
As shown in fig.~S11c~\cite{supplementary}, the dependence of the critical anisotropy parameter  $\beta_\textrm{c}$ on the $L_\textrm{H}/L_\textrm{D}$ ratio for different values of $\alpha$, with $h$ held constant at 0.35. 
Notably, for any given value of $\alpha$, there exists a finite range of $L_\textrm{H}/L_\textrm{D}$ where $0<\beta_\textrm{c}<1$.

In addition to the anisotropy parameters $\alpha$ and $\beta$, a key parameter in the model is the ratio $L_\textrm{H}/L_\textrm{D}$.
The emergence of the S-AL as the lowest energy phase within a specific range of applied magnetic fields requires particular values of all three parameters.
This is illustrated by the magnetic phase diagrams shown in Fig.~\ref{fig:MicroMag_intB}.

First, let us consider the isotropic case ($\alpha=\beta=1$) where S-AL phase does not emerge as a stable phase, Fig.~\ref{fig:MicroMag_intB}\textbf{a}.  
For low values of $L_\textrm{H}/L_\textrm{D}$, the system exhibits typical behavior of 2D exchange-frustrated magnets~\cite{Leonov_15}. 
In particular, as $L_\textrm{H}/L_\textrm{D} \rightarrow 0$, the cone-SS phase becomes the lowest energy state across the entire field range, from zero up to saturation at $h = 1$.  
With increasing DMI contributions (increasing $L_\textrm{H}/L_\textrm{D}$), the conical phase starts to compete with the cycloidal-SS.  
For $L_\textrm{H}/L_\textrm{D} \gtrsim 0.7$, the stability range of the cycloidal-SS phase extends beyond the saturation field of the cone-SS, $h>1$, effectively eliminating the conical phase.  
As the ratio increases further to $L_\textrm{H}/L_\textrm{D} > 0.85$, in the applied field, the system exhibits a first-order phase transition from a cycloidal-SS to a hexagonal SL, followed by another first-order transition to a saturated FM state. 
This behavior is typical of isotropic chiral magnets, where DMI competes with exchange frustration~\cite{Nandy_PRL2016}. 
Notably, the transition from the SL phase to the saturated state in such frustrated chiral magnets represents a first-order phase transition.
In systems with dominating DMI, isolated antiskyrmions remain stable only in a narrow range of fields~\cite{Kuchkin_20i}.
In these conditions, the S-AL phase obviously cannot be stable.

Figure~\ref{fig:MicroMag_intB}\textbf{b} shows the phase diagram for the case of anisotropic exchange interactions and DMI, with $\alpha = \beta = 0.1$.
In this scenario, the S-AL phase appears within the intermediate range of $0.24 \leq L_\textrm{H}/L_\textrm{D} \leq 0.56$.
Similar to the isotropic model discussed above, the lower and upper bounds of this range can be attributed to the effective transition of the system to the models of anisotropic exchange-frustrated magnet and anisotropic chiral magnet~\cite{Hoffmann-NatCommun2017,kuchkin_23monoaxial}, respectively.  
In the pure model of a chiral magnet{--}without exchange-frustration{--}even at relatively weak anisotropy in DMI, $\beta\lesssim 0.7$, the SL becomes unstable~\cite{Kuchkin_25}.
In our model, for $L_\textrm{H}/L_\textrm{D} \gtrsim 0.56$, the SL remains stable even for $\beta = 0.1$ exclusively due to the presence of exchange frustration.

The emergence of the S-AL phase as the lowest energy state can be explained by its distinct magnetization properties. 
The S-AL phase is characterized by magnetization modulations with the opposite helicity angle,
$$
\gamma = \text{arctan2}\!\big( \mathbf m \cdot \hat\varphi,\; \mathbf m \cdot \hat\rho \big),
$$
where $\hat\rho$ and $\hat\varphi$ denote the radial and azimuthal unit vectors in the film plane, respectively, along orthogonal directions (see Fig.~\ref{fig:MicroMag_intA}\textbf{c}).
When the DMI coupling strength is reduced along the weak axis ($y$-direction), the total DMI energy contribution remains nearly unchanged. 
This occurs because, along the $y$-axis, the helicity angle of cycloidal modulations in the skyrmion chain is $\gamma=0$ while along the chain of antiskyrmions $\gamma=\pi$.
These chains of skyrmion and antiskyrmions are well seen in Fig.~\ref{fig:MicroMag_intA}\textbf{b}.  
As a result, the DMI contribution to the energy of the S-AL phase tends to be independent of the DMI strength along the $y$-axis and is thus independent of $\beta$.  
In contrast, the DMI contribution to the energy of the SL phase is highly sensitive to DMI anisotropy. 
As $\beta$ decreases, the energy of the SL phase rises, while the energy of the S-AL phase remains unaffected.  
It is important to note that this mechanism is feasible only in the presence of sufficiently strong exchange frustration.
In systems where DMI dominates, the S-AL phase is inherently unstable. Conversely, in systems where exchange frustration dominates, the stable phases below the saturation field are limited to cone-SS and cycloidal-SS. 
Our analysis shows that the energy balance can shift in favor of the S-AL phase by introducing anisotropy ($\alpha$) into the exchange energy terms.  

In conclusion, the stability of the S-AL phase relies on three critical factors:  
I) DMI and exchange frustration must contribute comparably to the energy of SS ($L_\textrm{H}$ and $L_\textrm{D}$ are of the same order).  
II) The exchange interaction must exhibit anisotropy $\alpha\neq 1$.  
III) The DMI must also be anisotropic $\beta\neq 1$.  
Notably, our analysis also indicates that the stability of the S-AL phase requires correlated anisotropy in exchange and DMI, ensuring that the hard exchange axis coincides with the hard DMI axis ($\alpha< 1$ and $\beta< 1$). 
This alignment reflects the natural behavior of magnetic systems with broken symmetry, confirming that the micromagnetic model presented here corresponds to a realistic physical system. 

\begin{figure*}
\centering
\includegraphics[width=1.0\linewidth]{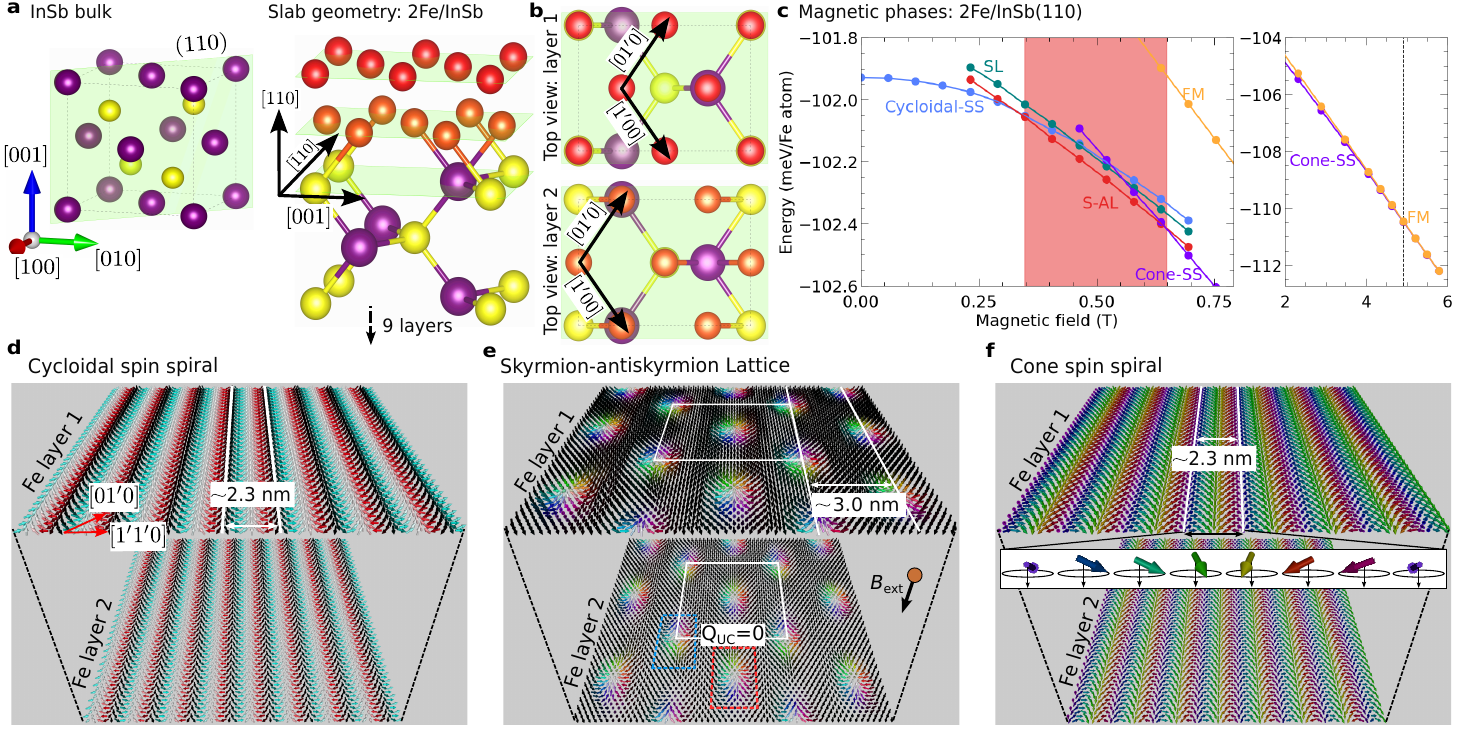}
\caption{\textbf{2Fe/InSb(110), a prototypical 2D magnetic heterostructure:} \textbf{a,} The slab geometry: a two-atomic-layer-thick Fe film grown on an InSb(110) substrate, forming a magnet/semiconductor heterostructure.
\textbf{b,} The thin magnetic layer on the surface of a semiconductor adopts a lattice structure with 2D crystallographic axes, [$1^\prime00$] and [$01^\prime0$]. 
\textbf{c,} Within atomistic lattice model~\eqref{spin_hamiltonian}, energy lines representing all competing phases are plotted against the external magnetic field $B_\textrm{ext}$, applied perpendicular to the heterostructure. 
The S-AL phase, highlighted by the red region, is the ground state within a specific range of $B_\textrm{ext}$. The first-order phase transitions occur at the boundaries of this range: from the cycloidal-SS to the S-AL at low fields, and from the S-AL to the cone-SS at higher fields. 
The rightmost panel further confirms the second-order phase transition between the cone-SS and FM phases, as evidenced by the merging of energy lines beyond the vertical line.
\textbf{d,} Spin configuration of the zero-field cycloidal-SS ground state, obtained within spin-lattice simulations. 
\textbf{e,} The S-AL phase at $B_\textrm{ext}$ = 0.5 Tesla. 
In this materials, the S-AL is also a charge (topological) neutral state due to equal number $Q=\pm 1$ topological charges. Similar to the micromagnetic model, the representing unit cell carries \textit{net-zero} topological charge $i.e.,$ $Q_\textrm{UC}=0$.
\textbf{f,} Cone-SS phase at $B_\textrm{ext}$ = 0.9 Tesla. To improve visibility, the identical magnetic states in both Fe layers are spatially separated. 
}
	\label{fig:real-material}
\end{figure*}

To gain a deeper understanding of the factors stabilizing the S-AL phase, we extended our 2D model to include the magnetocrystalline anisotropy energy term (see Methods for details).
As shown in fig.~S13~\cite{supplementary}, the easy-axis uniaxial anisotropy significantly favors the cone-SS and FM phases over the lattice phase.
Consequently, the critical field for the S-AL to cone-SS phase transition decreases with increasing uniaxial anisotropy, narrowing the magnetic field window for the S-AL phase.

While skyrmion-antiskyrmion pairs generally exhibit a strong tendency towards annihilation, the anisotropic model can stabilize metastable clusters of both even and odd numbers of skyrmions and antiskyrmions under finite magnetic fields.
The attractive interaction between skyrmions and antiskyrmions in these clusters is evident from the snapshots provided in fig.~S14~\cite{supplementary}.
It also demonstrates their stability over a wide range of applied magnetic fields.
More importantly, the skyrmions and antiskyrmions are found to form stable configurations without annihilation.
At low magnetic fields ($h = 0.5$), the cone-SS phase represents the global energy minimum, and all clusters in figs.~S14a-d~\cite{supplementary}, regardless of their composition, remain stable within the conical background magnetization.
In the pure model of a frustrated magnet, similar clusters embedded in the cone-SS, but consisting of a single type of particle, were previously discussed in Ref.~\cite{speight2020skyrmions}.
As the magnetic field increases beyond the critical field for the second-order phase transition to FM state ($h > 0.7$), the background magnetization becomes homogeneous, but all clusters remain intact (see figs.~S14e-h~\cite{supplementary}).

The above analysis of the micromagnetic model indicates that the criterion for the emergence of S-AL as the lowest energy state is met over a broad range of model parameters. Therefore, it is reasonable to expect the stable S-AL phase to appear in real systems.
As a proof of concept, in the following section, we present a 2D chiral magnet, identified through DFT calculations and spin-lattice simulations, as a promising candidate for the experimental observation of the phenomena described above.

\vspace{0.4cm}
\noindent\textbf {Potential real 2D magnet: DFT and atomistic spin-lattice simulations}

\noindent To demonstrate S-AL stability in real material, we present a two-atomic-layer-thick Fe film on an InSb(110) substrate, as depicted in Figs.~\ref{fig:real-material}\textbf{a} and \textbf{b}.
It is worth noting that we initially studied the 2Fe/InSb(110) system to explore magnetically ordered 2D structures on semiconducting substrates. Simulations based on DFT-derived parameters revealed the S-AL state as the lowest-energy configuration. This unexpected result prompted the development of a micromagnetic model that explains the stabilization mechanism and generalizes the findings beyond this specific material.
The zincblende InSb is a semiconductor that has a notably strong spin-orbit coupling~\cite{Silva_PRB1997} and a well-characterized (110) surface~\cite{Klijn_PRB2003}.
The surface unit cell is rectangular, and each layer is characterized by a distinct arrangement of In and Sb atoms.
The deposition of Fe films on InSb(110) surfaces can induce a significant interfacial DMI among magnetic Fe atoms. 
This DMI, stemming from SOC and broken inversion symmetry at the interface (Fig.~\ref{fig:real-material}\textbf{a}), is essential for stabilizing chiral magnetic states, especially in systems where frustrated exchange interactions primarily govern the SS state.

In contrast to traditional 2D chiral magnets, which often utilize heavy-metal substrates ~\cite{FerrianiPRL,Bode_Nature2007,Romming_13,Nandy_PRL2016,Dupe_NatCommun2014}, our 2D slab geometry integrates a binary semiconductor, leading to a distinctive interfacial configuration.
This design is inspired by the well-characterized 2Fe/GaAs(110) interface~\cite{Winking_APL2008, Grunebohm_PRB2009, Ifflander_PRL2015}, which provides a compelling rationale for our choice of a similar interface.
Furthermore, the (110) binary semiconductor surface, with its characteristic $C_{1v}$ symmetry~\cite{S-AL_Mat}, is essential for providing the anisotropic interaction parameter space as established in our micromagnetic model.
In the magnetic unit cell, each of the two magnetic layers (the surface and subsurface layers), consists of four Fe atoms. Each layer in the substrate, however, contains only two atoms, In and Sb. This arrangement yields inequivalent magnetic sites, strikingly different from the standard single-atom-per-layer ultrathin 2D magnets.

We have employed \textit{ab~initio} electronic structure calculations to characterize the relaxed interface of the 2Fe/InSb(110) slab geometry. 
As shown in Supplementary Table~S1, the magnetic moments of the Fe atoms in the 2Fe/InSb(110) sample are consistent with values previously reported for the 2Fe/GaAs(110) system~\cite{Grunebohm_PRB2009}. 
The relaxed magnetic slab is then used to calculate all material-specific atomistic interaction parameters, including exchange, DMI, and magnetocrystalline anisotropy.
For exchange and DMI, calculations extend beyond nearest neighbors to include several neighboring shells around each magnetic atom. 
This approach allows the parameters in the magnetic layers to capture both intra- and interlayer spin-spin interactions. Importantly, these atomistic parameters exhibit long-range behavior and display competing signs for each Fe atom.
Our first-principles simulation methodology is detailed in the Methods section, and DFT-based calculations of magnetic interaction parameters are presented in \textbf{Supplementary Note 3}~\cite{supplementary}.

The magnetic ground state is computed through full parameterization of the spin-lattice Hamiltonian~\eqref{spin_hamiltonian}, followed by atomistic spin dynamics simulations performed using the SPIRIT code~\cite{SPIRIT}.
%
%
For computational simplicity, we assume that all Fe atoms possess an identical average magnetic moment of $\mu_\textrm{s}=2.71~\mu_\textrm{B}$, a reasonable approximation.
Considering only Heisenberg exchange interactions, the exchange frustration driven magnetic solution is a SS state with a period of approximately 2.9 nm. This is detailed in \textbf{Supplementary Note 3~\cite{supplementary}}.

Upon inclusion of all interaction parameters, our atomistic simulations reveal a left-handed cycloidal-SS ground state with a period of 2.3 nm, as shown in Fig.~\ref{fig:real-material}\textbf{d}.
Notably, such atomic-scale SS states are often observed in ultrathin magnetic films grown on heavy-metal substrates ~\cite{Heinze_NatPhys2011,Nandy_PRL2016,Bode_Nature2007}.
In these systems, the interplay of exchange frustration and DMI typically governs the atomic-scale magnetic texture, with DMI dictating the specific axis of rotation. 
The presence of DMI often results in a slightly shorter period for the SS compared to that induced by exchange frustration alone.
To gain further insights into the system's behavior, we subject it to an external magnetic field, $B_\textrm{ext}$, oriented perpendicular to the magnetic film. 
This is modeled by adding a Zeeman energy term, $-\sum_i \mu_\textrm{s} \textbf{B}_\textrm{ext}\cdot \boldsymbol{\hat{m}}_i$, to our Hamiltonian~\eqref{spin_hamiltonian}.

Consistent with our micromagnetic model, we observed magnetic field-induced phase transitions, as presented in Fig.~\ref{fig:real-material}\textbf{c}. 
The cycloidal-SS state is the lowest energy magnetic configuration below the critical field, $B_\textrm{ext} \approx 0.35$ Tesla. 
Upon exceeding the critical field, a first-order phase transition results into a S-AL phase, as shown in Fig.~\ref{fig:real-material}\textbf{e}. 
A subsequent first-order phase transition transforms the S-AL phase into a cone-SS phase (see Fig.~\ref{fig:real-material}\textbf{f}) at a higher critical field, $\approx$ 0.65 Tesla.
A critical external magnetic field of approximately 5 Tesla results in a second-order phase transition to the saturated state.
Beyond this point, the energy lines corresponding to the cone-SS and saturated FM phases coalesce, as shown in the rightmost panel of Fig.~\ref{fig:real-material}\textbf{c}.
Remarkably, the 2Fe/InSb(110) system exhibits the same sequence of phase transitions as predicted by our micromagnetic model.
It is also noteworthy that in agreement with the micromagnetic simulations, both skyrmions and antiskyrmions exhibit an elongated shape.
These results collectively show that utilizing a semiconductor substrate to lower the symmetry is the crucial factor. This approach not only elucidates the stabilization mechanism but also proves its broad applicability to other material systems~\cite{S-AL_Mat}.

Consistent with the micromagnetic model, the ordinary SL phase is also stable in our spin-lattice model. As illustrated in Fig.~\ref{fig:real-material}\textbf{c}, this phase exhibits a higher energy compared to the S-AL phase.
The remarkable congruence between Fig.~\ref{fig:MicroMag_intA}\textbf{g} (micromagnetic simulations) and Fig.~\ref{fig:real-material}\textbf{c} (atomistic spin-lattice simulations), both unequivocally depicting the S-AL as the ground state, emphasizes the distinctive nature of chiral magnets with anisotropic interactions.

\vspace{0.4cm}
\noindent\textbf{Conclusions}

\noindent In conclusion, we have demonstrated a paradoxical phenomenon that has not been reported earlier. 
Specifically, we have demonstrated that an S-AL, composed of an equal number of skyrmions and antiskyrmions, becomes the lowest-energy state of 2D magnets under specific conditions. This phase is characterized by a \textit{net-zero} magnetic topological charge.
Since skyrmions and antiskyrmions possess opposite topological charges, one would expect them to annihilate.
However, we demonstrate that the S-AL stabilizes by the interplay of frustrated exchange and DMI in systems with interaction axes of both \textit{weak} and \textit{strong} character. This defines a distinct class of materials we term \textit{anisotropic frustrated chiral magnet}.
Additionally, our findings indicate that while uniaxial magnetocrystalline anisotropy is not a prerequisite for stabilizing the S-AL phase, it does favor the cone-SS and FM phases and consequently reduces the stability range of the S-AL phase.
Moreover, by combining DFT calculations with atomistic spin dynamics simulations, we propose 2Fe/InSb(110) as a realistic heterostructure for the experimental observation of this phenomenon. 
Our findings show that the S-AL in 2Fe/InSb(110) remains stable down to the atomic scale. 
Our micromagnetic simulations reveal a specific mechanism that stabilizes these magnetic phases in low-symmetry interfacial systems. This mechanism operates particularly well in a geometry like the (110) surface of semiconductors, due to its $C_{1v}$ symmetry.
%
This work not only presents the discovery of a new magnetic phase but also positions 2D chiral magnets with anisotropic interactions as promising materials for both fundamental research and practical applications.

\vspace{0.4cm}
\noindent\textbf{\large Methods}

\noindent\textbf{Micormagnetic simulations}\\
\noindent The micromagnetic model \eqref{Micro_Ham} was investigated using Mumax code~\cite{mumax}.
We use direct energy minimization starting with different initial spin configurations to find the equilibrium configurations.

To investigate the role of magnetocrystalline anisotropy, we added the following energy term in \eqref{Micro_Ham}:
$\mathcal{E}_a =  -Kn_z^2,$
where $K$ is the anisotropy constant, and the reduced anisotropy parameter is $u$ defined by $u=K/(M_\mathrm{s} B_\mathrm{c})$~\cite{reduced_ani}.
The sign of $u$ defines the easy-axis ($u>0$) or  easy-plane ($u<0$) anisotropy, respectively.
Figures~S13a and b~\cite{supplementary} illustrate the effect of magnetocrystalline anisotropy on the energy balance in the studied system.

\vspace{0.4 cm}
\textbf{Equilibrium state calculation:} In our micromagnetic calculations, the size of the simulated domain is $L_x \times L_y \times 1$ nm$^3$, along the $x$, $y$, and $z$ axes, respectively. 
To estimate the equilibrium energy density of different phases, we have utilized direct energy minimization.
To implement the higher-order exchange energy terms~\eqref{Micro_Ham_Ex}, we have used a custom effective field function built into Mumax. 
This approach had been used earlier in Refs.~\cite{Kuchkin_21, Kuchkin_23}. 
All calculations have been performed under periodic boundary conditions (PBC). 

First, we determined that the equilibrium propagation direction of the SS aligns with the ($\mathbf{e}_x$, $\mathbf{e}_y$, $0$) vector.  
To identify the equilibrium period of the SS state, we systematically varied the dimensions of the square domain, \( L_x = L_y \), which consisted of a \( 128 \times 128 \times 1 \) cuboids.  
The initial configurations were approximated by a homogeneous cycloidal-SS, with its period \( P \) chosen to be commensurate with the domain diagonal, satisfying the condition \( 2P = \sqrt{L_x^2 + L_y^2} \).  
For detailed insights into the initial implementation of the SS state, we refer the reader to the Mumax Script I.

The equilibrium SL and S-AL phases are determined by energy minimization of a rectangular unit cell, where skyrmion cores are strategically positioned at its center and four corners, as illustrated in the inset of figs.~S9a and b~\cite{supplementary}. 
This unit cell is embedded within a rectangular simulated domain, whose dimensions $L_x$ and $L_y$ can be tuned to control two key parameters: the \textit{core-to-core} distance $d=(L_x^2+L_y^2)^{1/2}$ along the diagonal and the angle $\theta = \textrm{tan}^{-1}(L_y/L_x)$ between the diagonal and the domain side. 
The latter parameter determines the shape of the lattices, specifying the degree of distortion from ideal hexagonal geometry with $\theta = 60^\circ$.

The equilibrium period of the exchange-frustrated SS is determined by the ratio of the second-order (\( \mathcal{A} \)) and fourth-order (\( \mathcal{B} \)) Heisenberg exchange terms, expressed as:  
$L_\textrm{H} = 4\pi \sqrt{{\mathcal{B}}/{|\mathcal{A}|}}$. 
This parameter plays a central role in our optimization scheme.  
The domain dimensions, \( L_x \) and \( L_y \), are directly related to \( L_\textrm{H} \) through a scaling factor.
For instance, the equilibrium unit cells of the SL and S-AL correspond to domains with mesh sizes of \( 71 \times 128 \times 1 \) and \( 56 \times 128 \times 1 \) cuboids, respectively. 
These meshes were chosen to maintain the ratio \( N_y / N_x \approx L_y / L_x \), where $N_x$ and $N_y$ are the number of cuboids along the $x$- and $y$-direction. 
This choice ensures that the cuboids are approximately square-shaped and that the numbers of cuboids per unit length along the $x$ and $y$ axes are nearly identical.
Accordingly, the optimal domain dimensions for SL are approximately \( L_x \approx 1.41 L_\textrm{H} \) and \( L_y \approx 2.54 L_\textrm{H} \), while for S-AL they are \( L_x \approx 1.3 L_\textrm{H} \) and \( L_y \approx 3 L_\textrm{H} \).  
In our simulations, with \( L_\textrm{H} = 50 \) nm, the domain volumes representing the equilibrium SL and S-AL unit cells are approximately \( 70.5 \times 122.1 \times 1 \) nm\(^3\) and \( 65 \times 150 \times 1 \) nm\(^3\), respectively.  

Micromagnetic simulations were carried out with the following parameters: exchange stiffness along the $x$-axis, $\mathcal{A}_x$ = $-10^{-17}$ J/m, and saturation magnetization, $M_s$ = 400 kA/m. 
The remaining interaction parameters in ~\eqref{Micro_Ham_Ex} and \eqref{Micro_Ham_DMI} can be determined using the parameters $\alpha$, $\beta$, $L_\textrm{H}$, and $L_\textrm{D}$. For details, we refer the reader to the Mumax Scripts.  

\vspace{0.4 cm}
\noindent\textbf{First-principles calculations}\\
\noindent Using \textit{ab initio} spin-polarized density functional theory (DFT) calculations within the Vienna Ab-initio Simulation Package (VASP)~\cite{hafner2008ab,kresse1996,Kresse}, we investigated the electronic and magnetic properties of the magnet/semiconductor heterostructure. 
In particular, this code is employed to determine the relaxed geometry of our slab construction.  
The asymmetric 2Fe/InSb(110) slab geometry, consisting of a bilayer Fe film on a 9-layer thick InSb(110) substrate, is constructed using the experimental lattice constant of bulk InSb, 6.479 \AA. A vacuum spacing of approximately 12 \AA~is maintained above and below the slab.
Projector-augmented wave (PAW) pseudopotentials~\cite{paw1, paw2} are used in conjunction with the Vosko-Wilk-Nusair (VWN) functional within the  local spin density approximation (LSDA)~\cite{vwn80} for exchange-correlation interactions.
A $16 \times 16 \times 1$ $\Gamma$-centered k-point mesh is considered for momentum-space integration over the two-dimensional Brillouin zone (2D-BZ). A plane-wave basis set with a cutoff energy of 500 eV is used for the expansion.
The atomic positions within the slab are relaxed until the forces on all atoms in the magnetic Fe layers (surface and subsurface) and the first three substrate layers adjacent to the Fe/InSb interface converge to a value below 0.001 eV/\AA.

\vspace{0.4 cm}
\textbf{Calculation of magnetic parameters:} Following geometry relaxation, all magnetic parameters are extracted within the DFT code JuKKR~\cite{KKR_Code}, which utilizes the full-potential Korringa-Kohn-Rostoker (KKR) Green's function method~\cite{Papanikolaou_2002, Bauer_thesis}. This method offers an exact description of the atomic cell shape~\cite{STEFANOU1990231, stefanou1991calculation}. 
The slab consists of 15 atomic layers (with 3 vacuum + 2 Fe layers + 7 InSb layers + 3 vacuum). This arrangement yields a vacuum spacing of approximately 7~\AA~on both the top and bottom surfaces of the slab. 
The momentum expansion of the Green's function has been truncated at $l_\textrm{max}=3$. The same exchange-correlation functional~\cite{vwn80} within the LSDA has been used.
The self-consistent calculations are performed using a 2D $k$-points grid of $40 \times 40\times 1$, with a contour integration involving 38 complex energy points in the upper half-plane, including 5 Matsubara poles. 
Self-consistent spin-polarized calculations, both with and without spin-orbit coupling, are performed to converge the unit cell potential.
With the converged potential, the pairwise Heisenberg exchange interactions and the DMI vectors were extracted using the infinitesimal rotation method\cite{LIECHTENSTEIN198765, Linch} with a k-mesh of a $200\times 200\times 1$.
We truncated exchange and DMI interactions at a cutoff radius of approximately 10~\AA, encompassing a total of 14 shells (7 intra-layer and 7 inter-layer shells) for each Fe atom.

\vspace{0.4 cm}
\noindent\textbf{Atomistic spin-lattice simulations}\\
\noindent To elucidate the magnetic configuration of our system, we have employed the computed material parameters: the Heisenberg exchange interactions ($\mathcal{J}_{ij}$), the DMI vector ($\textbf{D}_{ij}$), and the single-ion magnetocrystalline anisotropy ($\mathcal{K}$), as elaborated in \textbf{Supplementary Note 3~\cite{supplementary}}. 
These parameters serve as input for our extended Heisenberg model Hamiltonian, which takes the following form: 
\begin{align} 
\mathcal{H}&=\!-\!\sum_{i>j}\!\big[\!\mathcal{J}_{ij}\hat{\mathbf{m}}_i\!\cdot\!\hat{\mathbf{m}}_j+\mathbf{D}_{ij}\!\cdot\!(\hat{\mathbf{m}}_i\!\times\!\hat{\mathbf{m}}_j)\big]\!-\!\mathcal{K}\sum_i(\hat{\mathbf{m}}_i\!\cdot\!\hat{z})^2 \ , 
\label{spin_hamiltonian} 
\end{align} 
where $i$ and $j$ index the atomic sites within the domain, and $\hat{\mathbf{m}}$ denotes a unit vector along the magnetic moment direction.
As the model indicates, negative (positive) values of $\mathcal{J}$ correspond to antiferromagnetic (ferromagnetic) coupling, and their competition is crucial in determining the underlying ground state.
It is important to note that the large magnetic unit cell, comprising 8 Fe atoms and their long-range interactions, necessitates a comprehensive exploration of the parameter space.

Now, the magnetic state of the system is determined by numerically minimizing Eq.~\eqref{spin_hamiltonian} using the Monte Carlo (MC) method as implemented in SPIRIT~\cite{SPIRIT}, a GPU-accelerated atomistic code. 
To simulate the system with two magnetic layers, we have constructed 2D domains composed of $(80 \times 80)\times 2$ magnetic atoms along $x$ and $y$ directions, with periodic boundary conditions imposed in most cases.
To incorporate the effect of an external field, $B_\textrm{ext}$, applied perpendicularly to the magnetic domain, a Zeeman term, $-\mu_s\sum_i \textbf{B}_\textrm{ext} \cdot \hat{\mathbf{m}_i}$, is included in the model \eqref{spin_hamiltonian}.
To determine the zero-temperature ground state in the absence of $B_\textrm{ext}$, we utilize MC simulations by initializing the system with a random spin configuration at a high temperature of 100 K and then subsequently cooling it down to 0 K ($10^{-5}$ in our code). These calculations are performed under open boundary conditions (OBC) to promote the formation of a cycloidal-SS state. 
To further probe the equilibrium period, we impose an SS state with varying periods within a finite domain, followed by cooling the system from high temperatures.
The SS ground state serves as the initial configuration for subsequent finite magnetic field simulations. 
This state is heated to a temperature of T = 50 K under an applied magnetic field and subsequently cooled down in 2 K steps.
Importantly, regardless of the initial configuration, whether random or SS, the system invariably nucleates skyrmions and antiskyrmions at arbitrary positions within the domain under high temperature and magnetic field conditions.
For each temperature step, $10^5$ MC steps are used for thermalization, followed by additional $10^5$ steps to calculate physical quantities such as magnetization, energy, and topological charge.
Particular attention is paid to the lattice phases (SL and S-AL), which are carefully relaxed in 0.5 K steps below 30 K.
Additionally, the SL and S-AL phases are subjected to multiple heating and cooling cycles within a 20 K temperature range under both OBC and PBC conditions, allowing us to accurately determine their zero-temperature energies.



\vspace{0.25 cm}
\noindent\textbf{\large Conflict of Interest}\\
\noindent The authors declare no conflict of interest. 

\vspace{0.25 cm}
\noindent\textbf{\large Acknowledgements}\\
A.K.N., and S.B. acknowledge the support from the Department of Atomic Energy (DAE), Government of India, through the project Basic Research in Physical and Multidisciplinary Sciences via RIN4001. A.K.N. and S.B. acknowledge the computational resources, Kalinga cluster, at the National Institute of Science Education and Research (NISER), Bhubaneswar, India. A.K.N. thanks P. M. Oppeneer for the Swedish National Infrastructure for Computing (SNIC) facility. A.K.N. thanks Anamitra Mukherjee, Satyaprasad P Senanayak, Shovon Pal, Amaresh Jaiswal and Md Habib Ahsan for critical reading of the manuscript and stimulating discussions. The authors acknowledge discussions with Stefan Bl\"ugel. S.B. thanks Sandip Bera for fruitful discussions. N.S.K. acknowledges the European Research Council under the European Union's Horizon 2020 Research and Innovation Programme (Grant No.~856538 - project ``3D MAGiC'')
\vspace{0.25 cm}
\bibliography{Bib_Main}{}
\clearpage
\newpage

\begin{onecolumngrid}
\vspace{-2em}
\begin{center}	
	{
		\fontsize{12}{12}
		\selectfont
		\textbf{Supplemental material for: ``Paradoxical Topological Soliton Lattice in Anisotropic Frustrated Chiral Magnets''\\[5mm]}
	}
	
	\normalsize Sayan Banik\orcidA{}$^{1}$, Nikolai S. Kiselev\orcidB{}$^{2}$ and Ashis K. Nandy\orcidB{}$^{1}$\\
	{\small $^1$\textit{School of Physical Sciences, National Institute of Science Education and Research,\\ 
			An OCC of Homi Bhabha National Institute, Jatni 752050, India}\\
            $^2$\textit{Peter Gr\"unberg Institute, Forschungszentrum J\"ulich, 52425 J\"{u}lich, Germany}}
\end{center}	

\renewcommand{\thefigure}{\textbf{S\arabic{figure}}}
\noindent
\textbf{Supplementary Note 1$|$ Taylor expansion of spin-lattice Hamiltonian}

In this section, we present the derivations of the Hamiltonian describing a frustrated magnet in 2D.

\textbf{Heisenberg Exchange Interaction.}
Given the symmetry properties of the lattice in the Heisenberg model, the coupling constants $\mathcal{J}_{ij}$ between the $i$-th and $j$-th atoms can be categorized into distinct shells defined by the lattice symmetry. 
Each site $i$ and its neighboring sites $k$ and $m$ have coupling constants $\mathcal{J}_{ik}$ and $\mathcal{J}_{im}$ that are equivalent under the symmetry operations of the point group corresponding to the crystal. 
This symmetry allows the position vectors $\textbf{r}_{ik}$ and $\textbf{r}_{im}$ to transform into each other.\par
We define a complete set of lattice sites that are symmetry-equivalent to form a shell, with each shell being labeled by an integer $s$. 
The coupling constant associated with each shell is denoted as $J_s$. 
For instance, the first shell (nearest neighbors) corresponds to $s=1$, the second shell (next nearest neighbors) to $s=2$, and so forth.
Assuming that the vector $\mathbf{m}$ is a unit vector, the Heisenberg exchange interaction energy per one magnetic atom has the following form
\begin{align}
    \mathcal{H}_\mathrm{e}=-\sum_{i>j}^N\mathcal{J}_{ij}\mathbf{m}_i\cdot\mathbf{m}_j 
    =\sum_s^S\frac{1}{4}J_s\sum_{k,l,m}[\mathbf{m}(\mathbf{r})-\mathbf{m}(\mathbf{r}+a(k\mathbf{e}_x+l\mathbf{e}_y+m\mathbf{e}_y))]^2 
\label{spin-lattice_Exchnage}
\end{align}
where $\textbf{r}$ represents the position vector of the atom, and $k, l, m$ are sets of integer or half-integer indices that correspond to the shell $s$. 
We consider a crystal with a simple cubic Bravais lattice. 
Moving to the continuum limit, where the discrete spins $\mathbf{m}_i$ transit to a smoothly varying field $\mathbf{m}(\mathbf{r})$, we go beyond the conventional micromagnetic approximation and consider the higher-order terms in the series expansion.  
Let us consider a quasi-2D case, assuming that magnetization is homogeneous along the $z$-axis, $\mathbf{m}\equiv\mathbf{m}(x,y)$. 
In continuum approximation for cubic crystals, the Heisenberg exchange interaction with the terms up to fourth order can be written as~\cite{Rybakov_2022}:
\begin{align}
E_\mathrm{e}=&\int_{\mathbb{R}^2} \left\{\mathcal{A}\Biggl[\Biggl(\dfrac{\partial \mathbf{m}}{\partial x}\Biggr)^{\!\!2}+\Biggl(\dfrac{\partial \mathbf{m}}{\partial y}\Biggr)^{2}\Biggr]\right. 
\!+\left.\mathcal{B}\!\Biggl[\frac{\partial^2\mathbf{m}}{\partial x^2}-\frac{\partial^2\mathbf{m}}{\partial y^2}\Biggr]^{2}
\!+\mathcal{C}\Biggl[\frac{\partial^2\mathbf{m}}{\partial x \partial y}\Biggr]^{2}\right\}t\, \mathrm{d}x \, \mathrm{d}y
\label{Micro_Ham_Ex0_Method}
\end{align}
where $t$ is the plate thickness. The derivation provides linear relations between micromagnetic parameters $\mathcal{A}$, $\mathcal{B}$, $\mathcal{C}$ and exchange constants $J_s$, connecting the spin-lattice model to a continuum model with higher order terms.  
\begin{equation}
    \mathcal{A}=\frac{1}{a}\sum_{s}\textsf{a}_sJ_s; \mathcal{B}=-a\sum_{s}\textsf{b}_sJ_s; \mathcal{C}=-a\sum_{s}\textsf{c}_sJ_s 
\label{micro_latt}
\end{equation}
where $a$ is the lattice constant.
The positive coefficients $\textsf{a}_s$, $\textsf{b}_s$ and $\textsf{c}_s$ depend on the crystal lattice type. 
We express $\mathcal{A}$ and $\mathcal{B}$, $\mathcal{C}$ in J/m and J$\cdot$m units, respectively.

We now consider a simple cubic lattice with the lattice constant $a$ and coupling constants for the first four shells $\Tilde{J}_{1,2,3,4}$ which reproduce the same material parameters $\mathcal{A}$, $\mathcal{B}$, $\mathcal{C}$. 
According to~\eqref{micro_latt}, these constants must satisfy the following system of equations (see Ref.~\citenum{Rybakov_2022} for details),
\begin{align}
    \mathcal{A}&=\frac{1}{a}\Big(\frac{1}{2}\Tilde{J}_1+2\Tilde{J}_2+2\Tilde{J}_3+2\Tilde{J}_4\Big)\nonumber\\
    \mathcal{B}&=-a\Big(\frac{1}{96}\Tilde{J}_1+\frac{1}{24}\Tilde{J}_2+\frac{1}{24}\Tilde{J}_3+\frac{1}{6}\Tilde{J}_4\Big) \label{micro_to_latt}\\ 
    \mathcal{C}&=-a\Big(\frac{1}{48}\Tilde{J}_1+\frac{1}{3}\Tilde{J}_2+\frac{7}{12}\Tilde{J}_3+\frac{1}{3}\Tilde{J}_4\Big)\nonumber
\end{align}
The model can be simplified by considering only the neighbors sitting along orthogonal directions.
It means that we can exclude the interaction with other neighbors: $\Tilde{J}_2$ = 0, $\Tilde{J}_3$ = 0, and $\mathcal{C}=2\mathcal{B}$. 
We refer to that model as \textit{simplified effective model}. This model is depicted in Fig.~\textbf{1}, see the Main text, where the next after-nearest neighbor exchange $J_1$ ($J_2$) is $\Tilde{J}_1$ ($\Tilde{J}_4$)  in Eq.~\ref{micro_to_latt}.

Let us show that in the case of $\mathcal{C}=2\mathcal{B}$, the integral \eqref{Micro_Ham_Ex0_Method} can be reduced to the exchange energy term Eq.~(3) in the main text. 
For a smooth twice differentiable function $f$, one can prove the identity:
\begin{align}
2\left(\frac{\partial^2 f}{\partial x \partial y}\right)^{\!\!2}
-2\,\,\frac{\partial^2 f}{\partial x^2}\cdot\frac{\partial^2 f}{\partial y^2}=
\frac{\partial^2}{\partial y^2} \left[ \!\left( \frac{\partial f}{\partial x} \right)^{\!\!2} \right]
+\frac{\partial^2}{\partial x^2} \left[\! \left( \frac{\partial f}{\partial y} \right)^{\!\!2} \right]
-\frac{2\partial^2 }{\partial x \partial y} \left(\frac{\partial f}{\partial x} \cdot \frac{\partial f}{\partial y} \right) 
\!  
\label{identity}
\end{align}
It is easy to show that the integration of every term on the right-hand side of \eqref{identity} over $\mathbb{R}^2$ can be reduced to the boundary integral.
For instance, the first term in the right-hand side of \eqref{identity} can be written as
$$
\int_{\mathbb{R}^2} \frac{\partial^2}{\partial y^2} \left[ \left( \frac{\partial f}{\partial x} \right)^2 \right] \, \mathrm{d}x \, \mathrm{d}y = \int_{\partial \mathbb{R}^2} \frac{\partial}{\partial y} \left[ \left( \frac{\partial f}{\partial x} \right)^2 \right] \, \mathrm{d}x ,
$$
where $\partial \mathbb{R}^2$ denotes the region's boundary.
Without losing generality, in an extended system, when one can ignore the presence of edges and use periodic boundary conditions, such terms can always be set to zero.
Thereby, for the smooth function $f$ defined in the whole $\mathbb{R}^2$ space, the integral of \eqref{identity} can be set to zero
\begin{align}
\int_{\mathbb{R}^2} \left\{2\left(\frac{\partial^2 f}{\partial x \partial y}\right)^{\!\!2}
-2\,\,\frac{\partial^2 f}{\partial x^2}\cdot\frac{\partial^2 f}{\partial y^2}\right\}\, \, \mathrm{d}x \, \mathrm{d}y=0
\!  
\label{integral_identity}
\end{align}

Using the integral \eqref{integral_identity}, one can show that for $\mathcal{C}=2\mathcal{B}$ and assuming that every component of the magnetization vector field represents a continuous twice differentiable function, the term \eqref{Micro_Ham_Ex0_Method} can be written as 
\begin{align}
E = \int_{\mathbb{R}^2} \Biggl\{ 
\mathcal{A}\Biggl[\,\left(\dfrac{\partial \mathbf{m}}{\partial x}\right)^{2} \,+\,\, \left(\dfrac{\partial \mathbf{m}}{\partial y}\right)^{2}\,\Biggr]
 + \mathcal{B}\Biggl[\left(\dfrac{\partial^2 \mathbf{m}}{\partial x^2}\right)^{2} + \left(\dfrac{\partial^2 \mathbf{m}}{\partial y^2}\right)^{2}\Biggr] 
\Biggr\} t \, \mathrm{d}x \, \mathrm{d}y
\label{Micro_Ham_Ex1_Method}
\end{align}
Finally, assuming that the exchange stiffness constants are not identical for orthogonal directions, the integral \eqref{Micro_Ham_Ex1_Method} can be written as the exchange energy term (3) in the main text.

Now, let's analyze the energy density for a spin spiral (SS) in the case of the frustrated magnet with $\mathcal{A}<0$ and $\mathcal{B}>$ 0.
As follows from \eqref{micro_to_latt}, the condition for $\mathcal{A}<0$ is $\Tilde{J}_4<-\Tilde{J}_1/4$.
We will consider the solutions that, in the most general case, can be considered as conical spin spiral (cone-SS), Supplementary Fig.~\ref{fig:spin-spiral}.
In the absence of an external magnetic field and other potential energy terms as \textit{e.g.} magnetocrystalline anisotropy, the solution can be written in the form of a flat-SS, $\textbf{m}(\mathbf{r})=\left(
\cos(\mathbf{q}\cdot\mathbf{r}),\sin(\mathbf{q}\cdot\mathbf{r}),0
\right)$, where $\mathbf{r}=(x,y,z)$ is a position vector and $\mathbf{q}$ is the flat-SS wave vector.
The energy density of flat-SS propagating along three different crystallographic directions, [111], [110], and [100], is given by,
\begin{align}
  \mathcal{E}_\mathrm{s} = \mathcal{A}_sq^2+
    \begin{cases}
      \frac{2}{3}\mathcal{C}q^4  & \text{($\textbf{q}$ $\parallel$ [111])},\\
      (\mathcal{B}+\frac{1}{2}\mathcal{C})q^4 & \text{($\textbf{q}$ $\parallel$ [110])},\\
      4\mathcal{B}q^4 & \text{($\textbf{q}$ $\parallel$ [100])},\\
    \end{cases}       
\end{align}
and the equilibrium wave vectors and the cone angle of a spiral in case of non-zero applied fields are as follows
\begin{align}
    \mathbf{q}=\frac{1}{2}\sqrt{\frac{-\mathcal{A}}{\mathcal{C}}}(\pm\hat{\mathbf{e}}_x\pm\hat{\mathbf{e}}_y\pm\hat{\mathbf{e}}_z),\quad\theta=\arccos({\frac{4M_sB_\mathrm{ext}\mathcal{C}}{3\mathcal{A}^2}}),
    \label{111}
\end{align}
\begin{align}
    \mathbf{q}=\frac{1}{\sqrt{2}}\sqrt{\frac{-\mathcal{A}}{2\mathcal{B}+\mathcal{C}}}(\pm\hat{\mathbf{e}}_x\pm\hat{\mathbf{e}}_y),\quad\theta=\arccos({\frac{M_sB_\mathrm{ext}(2\mathcal{B}+\mathcal{C})}{\mathcal{A}^2}}),
    \label{110}
\end{align}
\begin{align}
    \mathbf{q}=\frac{1}{2}\sqrt{\frac{-\mathcal{A}}{2\mathcal{B}}}(\pm\hat{\mathbf{e}}_\gamma),\quad \gamma\in \{x,y,z\},\quad \theta=\arccos({\frac{8M_sB_\mathrm{ext}\mathcal{B}}{\mathcal{A}^2}}).
    \label{100}
\end{align}

\begin{figure*}[h]
\centering
\includegraphics[width=15
cm]{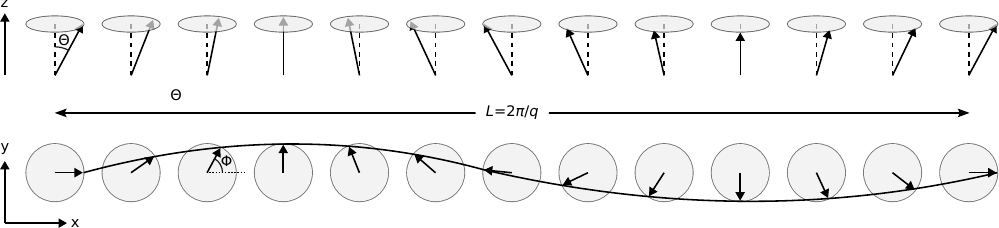}
\caption{~\small Conical SS with unit vector corresponding to each spin, 
$
\mathbf{n}(\mathbf{r})=\left[
\cos(\mathbf{q}\!\cdot\!\mathbf{r})\sin(\Theta),\sin(\mathbf{q}\!\cdot\!\mathbf{r})\sin(\Theta),\cos(\Theta)
\right]
$. 
The polar angle $\Theta$, measured with respect to the $z$-axis, remains fixed at each lattice site. The planar view in the bottom panel depicts a spin spiral with a period of $L=2\pi/q$, where the wavevector $\mathbf{q}$ is parallel to the $x$-axis. This cone-SS phase is subjected to an external magnetic field, $\mathbf{B}_\mathrm{ext}$, applied along the $z$-axis. The in-plane component makes an angle $\Phi$ ($\equiv \mathbf{q\cdot r}$) with the propagation direction. 
}
\label{fig:spin-spiral}
\end{figure*}

For the simplified effective model ($\mathcal{C} = 2\mathcal{B}$), the energy densities of the SS states with equilibrium periods along three different directions:
\begin{equation}
  \mathcal{E}_\mathrm{eq} =
    \begin{cases}
      -\dfrac{3\mathcal{A}^2}{16\mathcal{B}}  & \text{($\textbf{q}$ $\parallel$ [111])},\\[8pt]
      -\dfrac{\mathcal{A}^2}{8\mathcal{B}} & \text{($\textbf{q}$ $\parallel$ [110])},\\[8pt]
      -\dfrac{\mathcal{A}^2}{16\mathcal{B}} & \text{($\textbf{q}$ $\parallel$ [100])},\\[8pt]
    \end{cases}       
\end{equation}
Therefore, the energy comparison among the SSs with $\textbf{q}$-vectors along three different directions is as follows:
\begin{equation}
    \mathcal{E}_{[111]}<\mathcal{E}_{[110]}<\mathcal{E}_{[100]}
\end{equation}
and thus in 2D, the lowest energy state is an SS with $\textbf{q}\parallel$ [110].

Let's now consider the implementation of our simplified effective model in Mumax. 
We assume the following parameters are defined: saturation magnetization, $M_s$; exchange stiffness, $\mathcal{A}$; equilibrium period SS, $L_\mathrm{H}=2\pi/q$. 
The parameter $\mathcal{B}$ can be found from~\eqref{110} using $L_{\mathrm{H}}$:
\begin{equation}
    \mathcal{B}=-\frac{\mathcal{A}L_\mathrm{H}^2}{(4\pi)^2}
    \label{B}
\end{equation}
and the parameters $J_1$ = $\frac{\Tilde{J}_1}{a}$ and $J_4$ = $\frac{\Tilde{J}_4}{a}$ can be found from~\eqref{micro_to_latt},
\begin{equation}
    J_1=\frac{8(a^2\mathcal{A}+12\mathcal{B})}{3a^2}=\frac{8}{3}\mathcal{A}-\frac{2\mathcal{A}L_\mathrm{H}^2}{\pi^2a^2}=\frac{8}{3}\mathcal{A}-\frac{2}{\pi^2}\mathcal{A}\mathcal{N}^2
    \label{J1}
\end{equation}
\begin{align}
    J_4=-\frac{(a^2\mathcal{A}+48\mathcal{B})}{6a^2}&=-\frac{(a^2\mathcal{A}+3\mathcal{A}L_\mathrm{H}^2\pi^{-2})}{6a^2}\\ \nonumber
    &=-\frac{1}{6}\mathcal{A}+\frac{1}{2\pi^2}\mathcal{A}\mathcal{N}^2
    \label{J4}
\end{align}
where $\mathcal{N}$ = $L_\mathrm{H}/a$ represents mesh density, i.e., the number of cuboids per one period of the SS. 
The parameter $\mathcal{N}$ is the internal parameter of a finite difference scheme. 

From Eq.~\eqref{B}, the SS period can therefore be obtained as $L_\textrm{H}=4\pi\sqrt{\frac{\mathcal{B}}{-\mathcal{A}}}$.
For instance, with $\mathcal{A} = -10^{-17}$ J/m and $\mathcal{B} = 1.6\times10^{-34}$ J$\cdot$m, the equilibrium period of SS is $L_\mathrm{H} \sim$ 50 nm, and the energetically preferred direction for its propagation is [110].
Finally, we introduce anisotropy parameter in exchange interactions $\alpha$ such that the Eqs.~\eqref{micro_to_latt} reduces to,
\begin{align}
    &\mathcal{A}_x=\frac{1}{a}\Big(\frac{1}{2}\Tilde{J}_{1x}+2\Tilde{J}_{4x}\Big), \\
    \nonumber
    &\mathcal{A}_y=\frac{\mathcal{A}_x}{\alpha}=\frac{1}{a}\Big(\frac{1}{2}\Tilde{J}_{1y}+2\Tilde{J}_{4y}\Big),\\\nonumber
    &\mathcal{B}_x=-a\Big(\frac{1}{96}\Tilde{J}_{1x}+\frac{1}{6}\Tilde{J}_{4x}\Big), \\ 
    &\mathcal{B}_y=\frac{\mathcal{B}_x}{\alpha}=-a\Big(\frac{1}{96}\Tilde{J}_{1y}+\frac{1}{6}\Tilde{J}_{4y}\Big) \nonumber
    \label{micro_to_latt_ani} 
\end{align}

\textbf{Dzyaloshinskii-Moriya interaction.}
Next, we consider the contribution of the Dzyaloshinskii-Moriya Interaction (DMI) in a system characterized by $C_{nv}$ symmetry. We define the magnitude of the DMI coupling constant $|\mathbf{D}_{ij}|$, representing the interaction strength between the $i$-th and $j$-th atoms. As observed in the symmetric properties of exchange interactions, the DMI coupling constants between the $i$-th site and its neighboring sites $k$ and $m$, denoted as $|\mathbf{D}_{ik}|$ and $|\mathbf{D}_{im}|$, exhibit equivalent magnitudes due to symmetry considerations.

The DMI in the spin-lattice Hamiltonian can be written as follows:
\begin{equation}
    \mathcal{H}_\mathrm{D} = -\sum_{i>j}^N |\mathbf{D}_{ij}| \mathbf{d}_{ij} \cdot [\mathbf{n}_i \times \mathbf{n}_j] = \sum_{s}^S D_s \sum_{k,l,m} \mathbf{d}_{k,l,m} [\mathbf{n}(\mathbf{r}) \times \mathbf{n}(\mathbf{r}+a(k\mathbf{e}_x + l\mathbf{e}_y + m\mathbf{e}_z))]
    \label{spin-lattice_DMI}
\end{equation}
where $D_s$ represents the DMI coupling strength for $s$-th symmetry-defined shell. For consistency with the above, we consider only the first four shells.

Transitioning to the continuum limit ($\textbf{n}_i$ → $\textbf{n}(\textbf{r})$), the Hamiltonian~\eqref{spin-lattice_DMI} with accuracy up to third-order terms can be written as follows:
\begin{align}
    E_\mathrm{D} &= \int \mathcal{E}_\mathrm{D} \, d\mathbf{r} = \int \left( \mathcal{D}_1 \left(\Lambda^{(x)}_{xz} + \Lambda^{(y)}_{yz}\right) + \mathcal{D}_2 \left(\Lambda^{(xxx)}_{xz} + \Lambda^{(yyy)}_{yz}\right) + \mathcal{D}_3 \left(\Lambda^{(xyy)}_{xz} + \Lambda^{(xxy)}_{yz}\right) \right) d\mathbf{r}
    \label{Ham_DMI}
\end{align}
The Lifshitz invariants, which describe the spatial modulation of the magnetization due to DMI, are defined as follows:
\begin{align}
    \nonumber
    &\Lambda_{ij}^{(k)} = n_i \frac{\partial n_j}{\partial r_k} - n_j \frac{\partial n_i}{\partial r_k}, \\
    &\Lambda_{ij}^{(klm)} = n_i \frac{\partial}{\partial r_k} \frac{\partial}{\partial r_l} \frac{\partial n_j}{\partial r_m} - n_j \frac{\partial}{\partial r_k} \frac{\partial}{\partial r_l} \frac{\partial n_i}{\partial r_m}
\end{align}
The terms $\mathcal{D}_1$, $\mathcal{D}_2$, and $\mathcal{D}_3$ represent the constants of the first and third-order DMI terms, respectively, and can be related to the shell-based coupling constants of a simple cubic lattice through the relations:
\begin{align}
    \mathcal{D}_1 &= \frac{1}{a^2} \left({D}_1 + 2\sqrt{2} {D}_2 + \frac{4}{\sqrt{3}} {D}_3 + 2 {D}_4\right) \nonumber \\
    \mathcal{D}_2 &= \frac{1}{18} \left(3 {D}_1 + 6\sqrt{2} {D}_2 + 4\sqrt{3} {D}_3 + 24 {D}_4\right) \label{D123} \\
    \mathcal{D}_3 &= \frac{1}{6} \left(3\sqrt{2} {D}_2 + 4\sqrt{3} {D}_3\right)\nonumber
\end{align}
For consistency with the simplified effective model discussed above, we assume ${D}_2={D}_3=0$. Then, the above relations reduce to 
\begin{align}
\mathcal{D}_1 &= \frac{1}{a^2}\left( {D}_1  + 2{D}_4  \right), \nonumber \\
\mathcal{D}_2 &=  \frac{1}{6}{D}_1 +\frac{4}{3}{D}_4 , \label{simp_eff_mod_dmi} \\
\mathcal{D}_3 &= 0, \nonumber
\end{align}
and thus the third term in \eqref{Ham_DMI} can be omitted.
The constants $\mathcal{D}_1$ and $\mathcal{D}_2$ are expressed in units of [J/m$^{2}$] and [J], respectively.
We omit the DMI term with $\mathcal{D}_2$ in the following.
To justify the validity of such an approximation, let us consider the case where the energy density functional includes only the leading Heisenberg exchange energy term and the third-order DMI term:
\begin{equation} 
E=
\int\limits
\left(\mathcal{A} 
\sum_\alpha
\Bigg( \frac{\partial \mathbf{n}}{\partial r_\alpha} \Bigg)^{2}
+\mathcal{D}_2 
\left(
\Lambda_{xz}^{(xxx)}+\Lambda_{yz}^{(yyy)}
\right)
-M_\mathrm{s}\mathbf{B}_\mathrm{ext}\cdot \mathbf{n}\right)\,
d{\mathbf{r}}\, .
\label{Func-h2-dmi3}
\end{equation}
Assuming the external magnetic field is applied along the $y$-axis, the following equation defines the SS: 
\begin{equation} 
\mathbf{n}(\mathbf{r})=\left(
\cos(\mathbf{q}\cdot\mathbf{r})\sin(\Theta),\cos(\Theta),\sin(\mathbf{q}\cdot\mathbf{r})\sin(\Theta)
\right)
\label{cycloidal-SS}
\end{equation}
This SS corresponds to the scenario depicted in Fig.~\ref{fig:spin-spiral}, with the axes exchanged as follows: $\{x,y,z\}\rightarrow \{x,-z,y\}$.
According to \eqref{Func-h2-dmi3}, the energy density of such a configuration is
\begin{equation} 
\mathcal{E}=(\mathcal{D}_2q^3+\mathcal{A}q^2)\sin^2(\Theta)- B_\mathrm{ext}\cos(\Theta),
\label{e-h2-dmi3}
\end{equation}
It is seen that there is no equilibrium period of SS. 
It is important to emphasize that it is true irrespective of the sign of $\mathcal{A}$ and $\mathcal{D}_2$.
The global minimum of \eqref{e-h2-dmi3} is  $|q|\rightarrow \infty$ and the only metastable solution is $q=0$.
Since the actual values of $q$ are restricted between $-\pi/a$ and $+\pi/a$, where $a$ is the lattice constant, there is a critical value of $\mathcal{A}>\pi\mathcal{D}/a$ above which the global energy minimum of \eqref{Func-h2-dmi3} is a ferromagnet (FM), $i.e., q=0$.
When $\mathcal{A}<\pi\mathcal{D}/a$, the global energy minimum corresponds to the latest possible value of $|q|=\pi/a$.
Therefore, the only solutions to the model \eqref{Func-h2-dmi3} are either FM or an antiferromagnet.
Consequently, the third-order DMI term, in competition with the leading Heisenberg energy term, cannot stabilize the SS as the ground state.
The competition between high-order DMI and exchange may lead to a stable SS ground state only in the presence of a fourth-order Heisenberg exchange term.
In this case, the energy density of an SS is $\mathcal{E} \sim \mathcal{B}q^4 + \mathcal{D}_2q^3+\mathcal{A}q^2$, allowing for a potential energy minimum at a finite, nonzero $q$.
However, the contributions of higher-order terms, such as the fourth-order Heisenberg exchange and the third-order DMI, normally appear to be very weak compared to the other terms.
In conclusion, we assume that higher-order DMI terms can be omitted in the first approximation.

Focusing on the first-order DMI term exclusively, we derive the following:
\begin{align}    
    E_\mathrm{D} = \int \mathcal{E}_\mathrm{D} \, d\mathbf{r} = \int \left(\mathcal{D} \left(\Lambda^{(x)}_{xz} + \Lambda^{(y)}_{yz}\right)\right) d\mathbf{r}, \quad \text{where} \quad \mathcal{D} = \frac{D_1}{a^2}
    \label{1DMI}
\end{align}
In scenarios involving anisotropic systems, this expression simplifies to:
\begin{equation}
    E_\mathrm{D} = \int \mathcal{E}_\mathrm{D} \, d\mathbf{r} = \int \left(\mathcal{D}_x \Lambda^{(x)}_{xz} + \mathcal{D}_y \Lambda^{(y)}_{yz}\right) d\mathbf{r}
\end{equation}

\textbf{Isotropic model analysis.}
Taking into account the transformation of the Heisenberg energy term presented in the Method section for the case $\mathcal{C} = 2\mathcal{B}$, the functional that describes an isotropic magnetic system with frustrated exchange interaction and DMI is given by:
\begin{align}
    E(\mathbf{n})= \int\left(\mathcal{A}\Biggl[\,\left(\dfrac{\partial \mathbf{m}}{\partial x}\right)^{2} \,+\,\, \left(\dfrac{\partial \mathbf{m}}{\partial y}\right)^{2}\,\Biggr] 
 + \mathcal{B}\Biggl[\left(\dfrac{\partial^2 \mathbf{m}}{\partial x^2}\right)^{2} + \left(\dfrac{\partial^2 \mathbf{m}}{\partial y^2}\right)^{2}\Biggr] 
 +\mathcal{D}\Biggl[\Lambda^{(x)}_{xz}+\Lambda^{(y)}_{yz}\Biggr]-M_s\mathbf{B}_\mathrm{ext}\cdot\mathbf{n}\right)d\mathbf{r}
    \label{A_24_D1}
\end{align}
The energy density of a cycloidal-SS~\eqref{cycloidal-SS}, with the wave vector $q$ aligned along the [100] crystallographic direction, as derived from equation~\eqref{A_24_D1}, is 
\begin{equation}
    \mathcal{E}=(4\mathcal{B}q^4+\mathcal{A}q^2-\mathcal{D}q)\sin^2(\theta)-B_\mathrm{ext}\cos(\theta)
    \label{En_A_24_D1_100}
\end{equation}
The only real solution for equation~\eqref{En_A_24_D1_100} is given by:
\begin{equation}
    q=\frac{\sqrt[3]{6}\left(\sqrt{3}\sqrt{B^3(2A^3+27B)}+9B^2\right)^{2/3}-6^{2/3}AB}{12B\sqrt[3]{\left(\sqrt{3}\sqrt{B^3(2A^3+27B)}+9B^2\right)}}
\end{equation}
where $A = \dfrac{\mathcal{A}}{\mathcal{D}}$ and $B = \dfrac{\mathcal{B}}{\mathcal{D}}$. 
In the limiting case of $\mathcal{A}\xrightarrow{}0$, the solution simplifies to:
\begin{equation}
    q=\frac{1}{2^{4/3}}\sqrt[3]{\frac{\mathcal{D}}{\mathcal{B}}}, \quad L_\mathrm{D}=2\pi2^{4/3}\sqrt[3]{\frac{\mathcal{B}}{\mathcal{D}}}
    \label{DplusB}
\end{equation}
The energy density for a cycloidal-SS with $q$ aligned along [110] based on equation~\eqref{A_24_D1} is:
\begin{equation}
    \mathcal{E}=(2\mathcal{B}q^4+\mathcal{A}q^2-\mathcal{D}q)\sin^2(\theta)-B_\mathrm{ext}\cos(\theta)
    \label{En_A_24_D1_110}
\end{equation}
It is noted that the energy of the state described by equation~\eqref{En_A_24_D1_110} is lower than that of the SS in equation~\eqref{En_A_24_D1_100}. 
The corresponding wave vector solution is:
\begin{equation}
    q=\frac{\sqrt[3]{2}\left(\sqrt{3}\sqrt{B^3(4A^3+27B)}+9B^2\right)^{2/3}-2\sqrt[3]{3}AB}{2\cdot6^{2/3}B\sqrt[3]{\left(\sqrt{3}\sqrt{B^3(4A^3+27B)}+9B^2\right)}}
\end{equation}
As $\mathcal{A}\xrightarrow{}0$, this solution reduces to:
\begin{equation}
    q=\frac{1}{2\sqrt[3]{B}}=\frac{1}{2}\sqrt[3]{\frac{\mathcal{D}}{\mathcal{B}}}, \quad L_\mathrm{D}=4\pi\sqrt[3]{\frac{\mathcal{B}}{\mathcal{D}}}
    \label{LD}
\end{equation}
In the scenario where $\mathcal{D}\xrightarrow{}0$, the solution simplifies as referenced in equations~\eqref{110} and~\eqref{B}:
\begin{equation}
    q=\frac{1}{2}\sqrt{\frac{-\mathcal{A}}{\mathcal{B}}}, \quad L_\mathrm{H}=4\pi\sqrt{\frac{\mathcal{B}}{-\mathcal{A}}}
    \label{LH}
\end{equation}

Finally, introducing anisotropy into both the Heisenberg exchange and the DMI results in the following functional:
\begin{align}
    E(\mathbf{n})= \int\bigg[\biggl(\mathcal{A}_x\left(\dfrac{\partial \mathbf{m}}{\partial x}\right)^{2} + \mathcal{A}_y \left(\dfrac{\partial \mathbf{m}}{\partial y}\right)^{2}\, 
 + \mathcal{B}_x\left(\dfrac{\partial^2 \mathbf{m}}{\partial x^2}\right)^{2} + \mathcal{B}_y\left(\dfrac{\partial^2 \mathbf{m}}{\partial y^2}\right)^{2}\biggr) 
 +\biggl(\mathcal{D}_x\Lambda^{(x)}_{xz}+\mathcal{D}_y\Lambda^{(y)}_{yz}\biggr)
 -M_s\mathbf{B}_\mathrm{ext}\cdot\mathbf{n}\biggr]d\mathbf{r}
    \label{final}
\end{align}
We define the anisotropies in the frustrated exchange as $\alpha$ and in the DMI as $\beta$, such that $\dfrac{\mathcal{A}_y}{\mathcal{A}_x} = \dfrac{\mathcal{B}_y}{\mathcal{B}_x} = \alpha$ and $\dfrac{\mathcal{D}_y}{\mathcal{D}_x} = \beta$, respectively.
The energy density Eq.~\eqref{final} is implemented within Mumax.
\\
 
\vspace{1cm}

\noindent\textbf{Supplementary Note 2$|$ Limiting cases within our Model}
\vspace{0.2 cm}

\noindent Our model, described by Eq.~\eqref{final}, demonstrates remarkable generality. It not only predicts the existence of the unprecedented skyrmion-antiskyrmion lattice (S-AL) phase as the energetically preferred ground state but also encompasses a broad spectrum of established theoretical and experimental scenarios, achieved by systematically exploring specific conditions in the model parameters. 
\vspace{0.2cm}

\noindent\textbf{Case I: Isotropic frustrated magnet}\\
\noindent We first investigate the role of exchange frustration by considering a simplified model where the DMI term in~\eqref{final} is neglected ($\mathcal{D}_x=\mathcal{D}_y=0$). 
Frustration in the exchange interactions is introduced by setting $\mathcal{A}_x = \mathcal{A}_y = \mathcal{A} < 0$ and $\mathcal{B}_x = \mathcal{B}_y = \mathcal{B} > 0$.
In the absence of an external magnetic field, the model yields a cone-SS with an analytical period $L_\textrm{H} \approx 50$ nm, as discussed in the main text.
In this frustrated magnetic system, subjected to an external perpendicular field $h$, only two equilibrium phases are observed across the entire range of fields: the cone-SS phase, as depicted in Fig.~\ref{fig:spin-spiral}, and the saturated FM.
Figure~\ref{SM_Case_I}\textbf{a} presents the energy density of the cone-SS as a function of its period at $h=0.5$. 
The equilibrium period, determined to be approximately 50.1 nm, aligns well with the theoretical value. 
Notably, the period of the cone-SS remains independent of the external magnetic field.

\begin{figure*}[htbp]
    \centering    
    \includegraphics[width=0.95\linewidth]{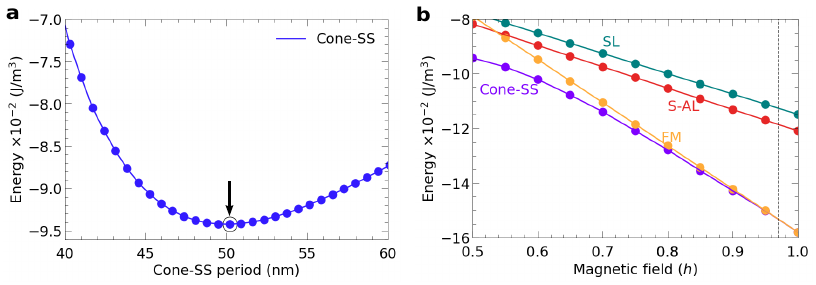}
    \caption{\textbf{Frustrated magnetic system with isotropic interactions:} \textbf{a}, Energy density plot as a function of the cone-SS period at perpendicular magnetic field $h=0.5$. The arrow indicates the numerical equilibrium period corresponding to the lowest energy of the cone-SS phase, $L_\textrm{H} \approx$ 50.1 nm. \textbf{b}, Energy density profiles as a function of $h$ for various magnetic phases: cone-SS, S-AL, SL, and FM. The vertical line indicates the second-order phase boundary between cone-SS and FM phases. The optimized hexagonal S-AL and SL phases remain metastable throughout the entire magnetic field range.}
    \label{SM_Case_I}
\end{figure*}

In Fig.~\ref{SM_Case_I}\textbf{b}, we present the energy density profiles for the cone-SS, FM, S-AL, and skyrmion lattice (SL) phases as a function of the external magnetic field $h$.
Two distinct regions, corresponding to the cone-SS and FM phases, are separated by a second-order phase boundary and identified as the energetically favored ground states across the entire magnetic field range.
In isotropic frustrated magnets, these two phases are commonly observed as ground states under specific magnetic field regions, as earlier demonstrated by the phase diagram in Ref.~\cite{Leonov_15}.
Our model, in the absence of magnetocrystalline anisotropy, accurately reproduces this limiting behavior.
Although the SL and S-AL possess equilibrium periods, they always occupy higher energy states than the cone-SS and saturated FM phases, irrespective of the applied magnetic field $h$.
Notably, even in the absence of DMI, the optimized S-AL exhibits lower energy than the optimized SL in the presence of exchange frustration.

\vspace{0.2cm}
\noindent\textbf{Case II: Conventional chiral magnet}\\
\noindent The second limiting case considers conventional chiral magnets with isotropic DMI and FM exchange interactions, as described by our model~\eqref{final} with $\mathcal{A}=\mathcal{A}_x=\mathcal{A}_y>0$ and $\mathcal{B}=0$. %
The DMI here lifts the degeneracy between clockwise and anti-clockwise SSs, resulting in the formation of a chiral phase.
The DMI term in model~\eqref{final} stabilizes a cycloidal-SS phase, consistent with the behavior observed in two-dimensional chiral magnets with interfacial DMI.
Without an external magnetic field, the ground state solution here is a right-handed cycloidal-SS, whose equilibrium period is exclusively governed by $4\pi\frac{\mathcal{A}}{|\mathcal{D}|}$~\cite{3DchiralMag}.
As a specific example, for $\mathcal{A}=10^{-17}$ J/m and $\mathcal{D}=12.5\times 10^{-19}$ J/m$^2$, the analytical period of the cycloidal-SS solution is approximately 100 nm.

A common scenario involves the transition of the cycloidal-SS state to a triangular lattice of magnetic solitons under an external magnetic field $h$ applied perpendicular to the 2D plate.
However, these lattices are composed of a single type of soliton: skyrmions for $\mathcal{D}_x=\mathcal{D}_y$ or antiskyrmions for $\mathcal{D}_x=-\mathcal{D}_y$~\cite{Hoffmann-NatCommun2017}.
Considering $\mathcal{D}_x=\mathcal{D}_y$, Fig.~\ref{SM_Case_II}\textbf{a} presents the energy density landscape as a function of $h$, revealing a stable SL phase bounded by two first-order phase transition lines at critical fields $h\sim 0.11$ and $h\sim 0.39$. 
The lattice is a regular hexagonal lattice of axisymmetric skyrmions.
Two critical field values define the phase boundaries separating the cycloidal-SS, SL, and saturated FM phases.
It is noteworthy that both the S-AL and antiskyrmion lattice phases are unstable in this DMI arrangement.

Additionally, we consider the case of anisotropic DMI, characterized by $\mathcal{D}_x \neq \mathcal{D}_y$ and controlled by the parameter $\beta$. 
In this anisotropic scenario, we identify the equilibrium SS period and hexagonal SL solution by systematically adjusting the domain size in our simulations.
The energy density profiles in Fig.~\ref{SM_Case_II}\textbf{b} reveal the presence of minima, which correspond to the system's minimum energy configuration for different $\beta$ values.
These simulations, conducted at zero external field ($h=0$), demonstrate the cycloidal-SS state as the ground state configuration.
Our analysis reveals a critical dependence of the SL phase on DMI anisotropy. Below a critical value of $\beta \approx 0.8$, the SL phase becomes energetically unfavorable and disappears. 
This critical behavior is evident in the domain size dependence: for domains exceeding the critical sizes (marked by red stars), the SL phase destabilizes and transitions into a stable SS solution.
This instability is analogous to the behavior observed in monoaxial chiral magnets, characterized by the absence of DMI along one direction, as described in Ref.~\cite{monoaxialCM}. 
In such systems, the energetically favored magnetic states are typically cycloidal-SS and saturated FM.
The critical behavior of $\beta$ in destabilizing the SL remains evident, even in the presence of a finite external field ($h=0.35$), as shown in Fig.~\ref{SM_Case_II}\textbf{c}.
%

\begin{figure*}[htbp]
    \centering    
    \includegraphics[width=1.0\linewidth]{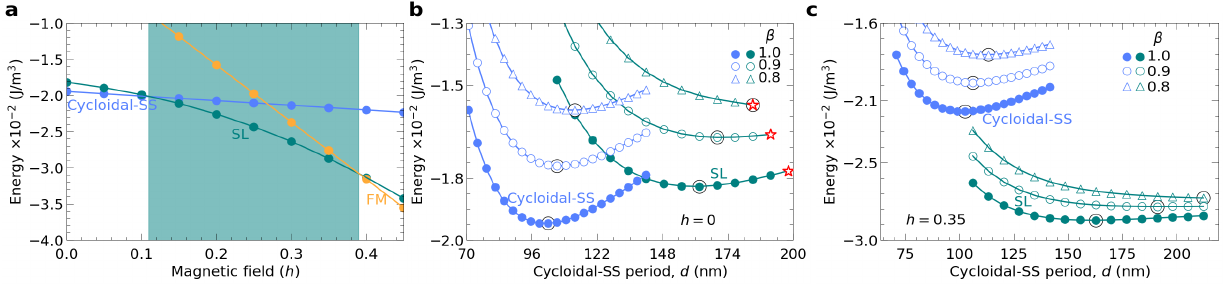}
    \caption{\textbf{Magnetic phases in conventional chiral magnets.} \textbf{a}, Energy density lines of cycloidal-SS, hexagonal SL, and saturated FM as a function of magnetic field $h$. The colored region, bounded by the two first-order phase transition lines, represents the equilibrium SL phase. This phase exists between the cycloidal-SS and saturated FM phases. \textbf{b}, at $h=0$, and \textbf{c}, at $h=0.35$, depict energy density variations with respect to the SS period and \textit{core-to-core} distances $d$ between two skyrmions for different DMI anisotropy parameter, $\beta$. The plot identifies regions of minimal energy (black circles) and points of instability for the SL (red asterisks). 
    }
    \label{SM_Case_II}
\end{figure*}

\vspace{0.2cm}

\noindent \textbf{Case III: Isotropic frustrated chiral magnet.}\\
\noindent There exist isotropic systems, e.g., ultrathin chiral magnets, which can exhibit both frustrated exchange and DMI interactions. This represents the third limiting case within our model.
The corresponding phase diagram, obtained within our model by setting $\alpha$ = $\beta$ = 1, is already presented in Fig.~\textbf{3a} of the main text.
However, the stability of the SL phase as the ground state is critically dependent upon a delicate balance between exchange and DMI energies. This energy balance can be controlled by the ratio $L_\textrm{H}/L_\textrm{D}$.
When the DMI strength becomes sufficiently strong compared to the exchange energy, we observe a first-order phase transition. In this transition, the cycloidal-SS transforms into the hexagonal SL phase.
In this limiting case, exchange frustration often stabilizes the SS state in such systems, but without a preferred rotational sense.   
The DMI lifts this degeneracy, selecting a specific rotational direction{--}the cycloidal-SS phase.
To model this isotropic case, we set the exchange interactions to $\mathcal{A}=\mathcal{A}_x=\mathcal{A}_y<0$ and $\mathcal{B}=\mathcal{B}_x=\mathcal{B}_y>0$, and the DMI strength to $|\mathcal{D}|=|\mathcal{D}_x|=|\mathcal{D}_y|$.
By tuning the DMI strength to a high value ($L_\textrm{H}/L_\textrm{D} = 1.2$), we observe a typical phase sequence of cycloidal-SS, hexagonal SL, and saturated FM states as the magnetic field $h$ is increased, as shown in Fig.~\ref{SM_Case_III}.
Within the field range from $h=2.4$ to $h=2.7$, the system exhibits the SL phase.
\begin{figure*}
    \centering    
    \includegraphics[width=0.4\linewidth]{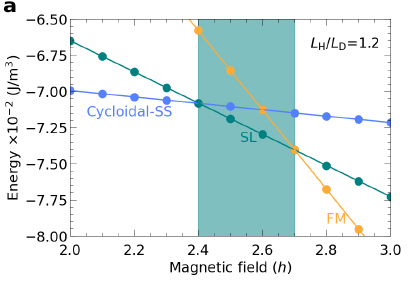}
	\caption{\textbf{Energy density profile for isotropic frustrated chiral magnet.} \textbf{a,} Energetically favored three phases: cycloidal-SS, hexagonal SL, and saturated FM. The SL phase is bounded by two first-order phase transition critical fields: $h=2.4$ (SS to SL) and $h=2.7$ (SL to FM).}
    \label{SM_Case_III}
\end{figure*}
As demonstrated in previous studies~\cite{Nandy_PRL2016,Heinze_NatPhys2011,Dupe-NatCommun2016}, atomistic spin-lattice simulations can also accurately capture the general behavior of such isotropic chiral magnets and other magnetic phases, including the SL phase.

\vspace{1cm}
\noindent\textbf{Supplementary Note 3$|$ 2Fe/InSb(110) a detailed analysis}\\
\vspace{0.1cm}

\noindent\textbf{Magnetic characterization of the 2D film within $\textbf{ab~initio}$ electronic structure calculations.}

\noindent The relaxed film geometry, obtained through structural optimization, serves as the basis for subsequent $ab~initio$ calculations.
The magnetic moments of each Fe atom are tabulated in Table~\ref{MagMom_tab1}. 
The initial four values pertain to the top Fe layer atoms, whereas the subsequent four are associated with the bottom Fe layer, as indicated by the numbering scheme in Figs.~\ref{atomistic_interactions}\textbf{a} and \textbf{b}.
To ensure consistency, we present values obtained from both VASP and KKR calculations. While generally consistent, a slight difference is noticeable, with KKR values typically exceeding those calculated using VASP.
As described in the main text, the atomistic spin-lattice simulations employed an average magnetic moment of 2.71 $\mu_\textrm{B}$ per Fe atom and an out-of-plane magnetocrystalline anisotropy of 0.6 meV per Fe atom, values obtained from KKR calculations.

\renewcommand{\thetable}{\textbf{S\arabic{table}}}
\begin{table}[h]
	\centering
	\begin{tabular}{|c|c|c|c|}\hline
	Fe atom& Magnetic mom. in $\mu_B$&Average ($\mu_B$)&$\mathcal{K}$ (meV/Fe atom) \\ 
     & KKR (VASP) & KKR (VASP)&\\ \hline
        Fe 1&2.76 (2.69)&& \\ \cline{1-2}
        Fe 2&2.80 (2.76)&&  \\ \cline{1-2}
        Fe 3&3.07 (2.92)&&  \\ \cline{1-2}
        Fe 4&3.06 (2.90)&2.71 (2.61)&0.6  \\ \cline{1-2}
        Fe 5&2.46 (2.42)&&  \\ \cline{1-2}
        Fe 6&2.41 (2.33)&&  \\ \cline{1-2}
        Fe 7&2.65 (2.49)&&  \\ \cline{1-2}
        Fe 8&2.50 (2.39)&&  \\ \hline
        \end{tabular}
	\caption{\textbf{Magnetic moments of Fe atoms and uniaxial magnetocrystalline anisotropy.} Each Fe atom has a different magnetic moment, which is a consequence of the anisotropic interactions present in our system. The magnetic moments are expressed in the units of $\mu_B$ while $\mathcal{K}$ is the out-of-plane uniaxial magnetocrystalline anisotropy.}
	\label{MagMom_tab1}
\end{table}

The magnetic heterostructure, featuring two distinct magnetic layers (Fig.~\ref{atomistic_interactions} and also in Figs.~\textbf{4a} and \textbf{b} in the main text), exhibits a rich complexity arising from eight unique Fe atomic configurations with diverse local environments.
This structural uniqueness, arising from low symmetry, results in significant variations in exchange and DMI parameters. The corresponding interaction parameters for each configuration, determined through KKR calculations, are tabulated in Tables~\ref{Fe1}-\ref{Fe8}.
It is crucial to note that the intricate nature of these interfacial systems results in a remarkably large parameter space. Furthermore, long-range interactions between magnetic atoms significantly influence their frustrated behavior.
\begin{figure*}[htbp]
    \centering
    \includegraphics[width=0.8\linewidth]{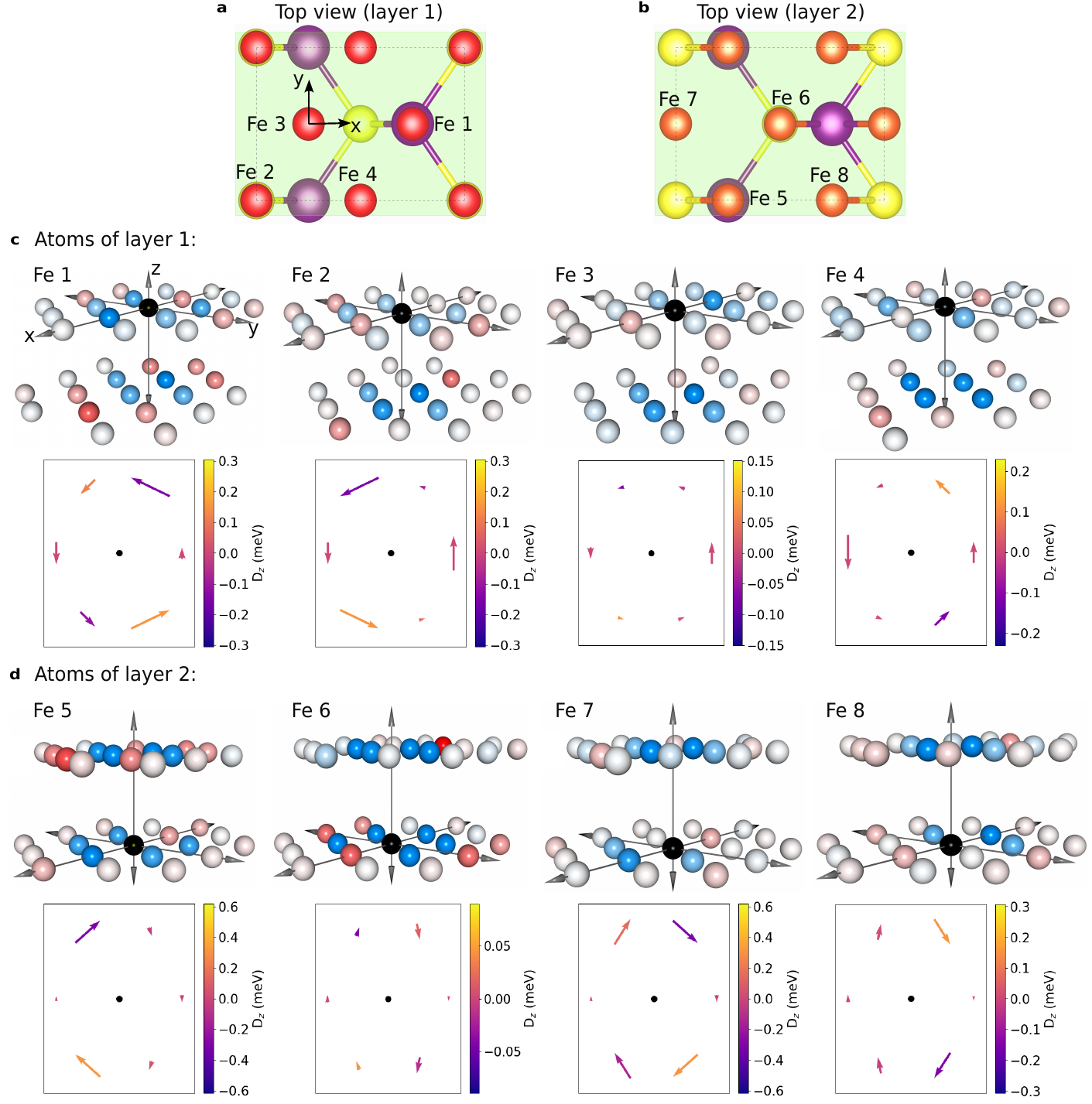}
    \caption{\textbf{Inequivalent Fe atoms in the heterostructure and corresponding magnetic interactions.} Top views of the two magnetic layers, each consisting of four Fe atoms: layer 1 in \textbf{a} and layer 2 in \textbf{b}. The distinct local environments of the eight Fe atoms, arising from the interfacial structure, result in anisotropic Fe$-$Fe interactions. For instance, considering the adjacent InSb(110) semiconductor layer, Fe3 exhibits a pronounced asymmetry in its nearest-neighbor coordination, with a clear absence of atoms in the $-x$ direction compared to the $+x$ direction. Furthermore, in atoms are present in the $\pm y$ directions, while the $\pm x$ directions lack In atoms. Similar asymmetric coordination patterns are observed for other Fe atoms. This broken symmetry at their coordination manifests in the material interaction parameters (exchange and DMI), leading to a pronounced anisotropic behavior, as detailed in panels \textbf{c,} and \textbf{d}.}
    \label{atomistic_interactions}
\end{figure*}
For further clarity, Figs.~\ref{atomistic_interactions}\textbf{c} and \textbf{d} (upper panels) provide a detailed visualization of the exchange coupling strengths between Fe atoms within the first and second layers, respectively.
Each sphere is color-coded according to the magnitude of exchange coupling strengths ($\mathcal{J}$), with more saturated hues denoting stronger interactions.
Reference Fe atoms are highlighted in black, with neighboring atoms color-coded to indicate the nature of the exchange interaction: blue for FM and red for antiferromagnetic. This visualization comprehensively maps both intra- and inter-layer couplings across the entire system.
For a detailed visualization of the DMI orientation for the first two neighbors of each Fe atom, we present vector plots in the lower panels of Figs.~\ref{atomistic_interactions}\textbf{c} and \textbf{d}. 
The color bar at right in these plots represents the strength of the DMI's $z$-component, which is smaller than the other components. 
This is non-negligible due to the surface roughness and the anisotropic substrate environment.
Importantly, these visualizations emphasize the presence of anisotropy in both exchange and DMI.
%
\begin{table}[ht]
\centering
\begin{tabular}{|l|c|c|c|c|c|c|c|}
\hline
\multicolumn{8}{|c|}{Interactions for atom: Fe 1} \\
\hline
& Neighbor atom (Fe) & $J_{ij}$ (meV)& $D_x$ (meV)& $D_y$ (meV)& $D_z$ (meV)& $|\textbf{D}|$ (meV)& $|\textbf{R}|$ (a) \\
\hline
Intra-layer& 2 & 6.097 & 1.230 & 0.397 & 0.305 & 1.327 & 0.433 \\
interactions& 2 & 6.097 & -1.230 & 0.397 & -0.305 & 1.327 & 0.433 \\
& 4 & 9.975 & 0.471 & -0.313 & -0.229 & 0.610 & 0.439 \\
& 4 & 9.975 & -0.471 & -0.313 & 0.229 & 0.610 & 0.439 \\
& 3 & 18.235 & 0.000 & 0.249 & -0.000 & 0.249 & 0.507 \\
& 3 & -4.662 & 0.000 & -0.473 & 0.000 & 0.473 & 0.507 \\
& 1 & -3.532 & -0.269 & -0.000 & -0.102 & 0.288 & 0.707 \\
& 1 & -3.532 & 0.269 & 0.000 & 0.102 & 0.288 & 0.707 \\
& 2 & 1.579 & -0.009 & 0.055 & -0.081 & 0.098 & 0.829 \\
& 2 & 1.579 & 0.009 & 0.055 & 0.081 & 0.098 & 0.829 \\
& 4 & 1.168 & 0.070 & 0.215 & -0.058 & 0.233 & 0.832 \\
& 4 & 1.168 & -0.070 & 0.215 & 0.058 & 0.233 & 0.832 \\
& 3 & -1.915 & -0.132 & -0.055 & 0.097 & 0.172 & 0.870 \\
& 3 & -1.915 & 0.132 & -0.055 & -0.097 & 0.172 & 0.870 \\
& 3 & 0.512 & -0.140 & -0.126 & 0.214 & 0.286 & 0.870 \\
& 3 & 0.512 & 0.140 & -0.126 & -0.214 & 0.286 & 0.870 \\
& 1 & 0.074 & 0.000 & -0.034 & -0.000 & 0.034 & 1.000 \\
& 1 & 0.074 & -0.000 & 0.034 & 0.000 & 0.034 & 1.000 \\
\hline
Inter-layer& 6 & 9.022 & -0.000 & -0.323 & -0.000 & 0.323 & 0.284 \\
interactions& 7 & 29.892 & -0.000 & 0.628 & 0.000 & 0.628 & 0.353 \\
& 8 & 8.547 & 0.067 & 0.349 & 0.345 & 0.496 & 0.433 \\
& 8 & 8.547 & -0.067 & 0.349 & -0.345 & 0.496 & 0.433 \\
& 5 & -4.370 & -0.124 & 0.119 & 0.163 & 0.237 & 0.639 \\
& 5 & -4.370 & 0.124 & 0.119 & -0.163 & 0.237 & 0.639 \\
& 5 & -6.111 & 0.186 & -0.071 & 0.042 & 0.204 & 0.639 \\
& 5 & -6.111 & -0.186 & -0.071 & -0.042 & 0.204 & 0.639 \\
& 6 & 0.202 & 0.029 & 0.047 & 0.098 & 0.113 & 0.762 \\
& 6 & -9.165 & -0.000 & -0.016 & 0.000 & 0.016 & 0.762 \\
& 6 & 0.202 & -0.029 & 0.047 & -0.098 & 0.113 & 0.762 \\
& 7 & -4.700 & -0.000 & -0.033 & -0.000 & 0.033 & 0.790 \\
& 7 & 0.378 & -0.204 & 0.185 & -0.346 & 0.442 & 0.790 \\
& 7 & 0.378 & 0.204 & 0.185 & 0.346 & 0.442 & 0.790 \\
& 6 & -0.878 & 0.013 & 0.100 & 0.127 & 0.163 & 1.040 \\
& 6 & -0.878 & -0.013 & 0.100 & -0.127 & 0.163 & 1.040 \\
& 8 & 0.438 & 0.124 & 0.076 & -0.163 & 0.218 & 1.090 \\
& 8 & 0.438 & -0.124 & 0.076 & 0.163 & 0.218 & 1.090 \\
\hline
\end{tabular}
\caption{\textbf{Heisenberg exchange interaction and DMI parameters for Fe 1 atom.} Both intra- and inter-layer interactions are taken into account. The distance between the neighbors is given in the units of the experimental lattice constant of InSb, 6.479 \AA.}
\label{Fe1}
\end{table}
\begin{table}[ht]
\centering
\begin{tabular}{|l|c|c|c|c|c|c|c|}
\hline
\multicolumn{8}{|c|}{Interactions for atom: Fe 2} \\
\hline
& Neighbor atom (Fe) & $J_{ij}$ (meV)& $D_x$ (meV)& $D_y$ (meV)& $D_z$ (meV)& $|\textbf{D}|$ (meV)& $|\textbf{R}|$ (a) \\
\hline
Intra-layer& 1 & 6.097 & -1.230 & -0.397 & -0.305 & 1.327 & 0.433 \\
interactions& 1 & 6.097 & 1.230 & -0.397 & 0.305 & 1.327 & 0.433 \\
& 3 & 7.625 & 0.208 & 0.040 & 0.150 & 0.259 & 0.444 \\
& 3 & 7.625 & -0.208 & 0.040 & -0.150 & 0.259 & 0.444 \\
& 4 & -0.545 & -0.000 & -0.461 & -0.000 & 0.461 & 0.508 \\
& 4 & -4.880 & 0.000 & 0.746 & -0.000 & 0.746 & 0.508 \\
& 2 & -5.217 & -0.245 & 0.000 & -0.324 & 0.406 & 0.707 \\
& 2 & -5.217 & 0.245 & -0.000 & 0.324 & 0.406 & 0.707 \\
& 1 & 1.579 & -0.009 & -0.055 & -0.081 & 0.098 & 0.829 \\
& 1 & 1.579 & 0.009 & -0.055 & 0.081 & 0.098 & 0.829 \\
& 3 & -0.341 & -0.106 & -0.151 & 0.018 & 0.185 & 0.835 \\
& 3 & -0.341 & 0.106 & -0.151 & -0.018 & 0.185 & 0.835 \\
& 4 & -2.272 & -0.039 & 0.074 & -0.058 & 0.102 & 0.871 \\
& 4 & 0.140 & 0.070 & 0.117 & -0.081 & 0.158 & 0.871 \\
& 4 & 0.140 & -0.070 & 0.117 & 0.081 & 0.158 & 0.871 \\
& 4 & -2.272 & 0.039 & 0.074 & 0.058 & 0.102 & 0.871 \\
& 2 & -1.321 & -0.000 & 0.356 & 0.000 & 0.356 & 1.000 \\
& 2 & -1.321 & 0.000 & -0.356 & -0.000 & 0.356 & 1.000 \\
\hline
Inter-layer& 5 & 16.590 & 0.000 & 0.131 & -0.000 & 0.131 & 0.300 \\
interactions& 8 & 39.792 & -0.000 & 0.408 & 0.000 & 0.408 & 0.342 \\
& 7 & 10.437 & -0.248 & 0.037 & -0.037 & 0.253 & 0.424 \\
& 7 & 10.437 & 0.248 & 0.037 & 0.037 & 0.253 & 0.424 \\
& 6 & 0.354 & 0.111 & -0.298 & -0.056 & 0.323 & 0.624 \\
& 6 & 0.354 & -0.111 & -0.298 & 0.056 & 0.323 & 0.624 \\
& 6 & -0.438 & 0.044 & -0.027 & -0.083 & 0.098 & 0.624 \\
& 6 & -0.438 & -0.044 & -0.027 & 0.083 & 0.098 & 0.624 \\
& 5 & 0.688 & -0.040 & -0.015 & 0.030 & 0.052 & 0.768 \\
& 5 & 0.688 & 0.040 & -0.015 & -0.030 & 0.052 & 0.768 \\
& 5 & -9.858 & 0.000 & -0.265 & -0.000 & 0.265 & 0.768 \\
& 8 & -6.573 & 0.000 & 0.109 & 0.000 & 0.109 & 0.785 \\
& 8 & -0.703 & -0.185 & -0.104 & -0.012 & 0.213 & 0.785 \\
& 8 & -0.703 & 0.185 & -0.104 & 0.012 & 0.213 & 0.785 \\
& 5 & -0.484 & 0.013 & 0.097 & 0.039 & 0.105 & 1.044 \\
& 5 & -0.484 & -0.013 & 0.097 & -0.039 & 0.105 & 1.044 \\
& 7 & 0.486 & 0.019 & 0.086 & -0.113 & 0.144 & 1.086 \\
& 7 & 0.486 & -0.019 & 0.086 & 0.113 & 0.144 & 1.086 \\
\hline
\end{tabular}
\caption{\textbf{Heisenberg exchange interaction and DMI parameters for Fe 2 atom.} Both intra- and inter-layer interactions are taken into account. The distance between the neighbors is given in the units of the experimental lattice constant of InSb, 6.479 \AA.}
\label{Fe2}
\end{table}

\begin{table}[ht]
\centering
\begin{tabular}{|l|c|c|c|c|c|c|c|}
\hline
\multicolumn{8}{|c|}{Interactions for atom: Fe 3} \\
\hline
& Neighbor atom (Fe) & $J_{ij}$ (meV)& $D_x$ (meV)& $D_y$ (meV)& $D_z$ (meV)& $|\textbf{D}|$ (meV)& $|\textbf{R}|$ (a) \\
\hline
Intra-layer & 2 & 7.625 & 0.208 & -0.040 & 0.150 & 0.259 & 0.444 \\
interactions& 2 & 7.625 & -0.208 & -0.040 & -0.150 & 0.259 & 0.444 \\
& 1 & 18.235 & -0.000 & -0.249 & 0.000 & 0.249 & 0.507 \\
& 1 & -4.662 & -0.000 & 0.473 & -0.000 & 0.473 & 0.507 \\
& 4 & 4.395 & 0.207 & 0.035 & -0.259 & 0.334 & 0.612 \\
& 4 & 2.537 & -0.214 & 0.057 & -0.028 & 0.223 & 0.612 \\
& 4 & 2.537 & 0.214 & 0.057 & 0.028 & 0.223 & 0.612 \\
& 4 & 4.395 & -0.207 & 0.035 & 0.259 & 0.334 & 0.612 \\
& 3 & -0.103 & -0.357 & 0.000 & 0.356 & 0.505 & 0.707 \\
& 3 & -0.103 & 0.357 & -0.000 & -0.356 & 0.505 & 0.707 \\
& 2 & -0.341 & 0.106 & 0.151 & -0.018 & 0.185 & 0.835 \\
& 2 & -0.341 & -0.106 & 0.151 & 0.018 & 0.185 & 0.835 \\
& 1 & 0.512 & 0.140 & 0.126 & -0.214 & 0.286 & 0.870 \\
& 1 & 0.512 & -0.140 & 0.126 & 0.214 & 0.286 & 0.870 \\
& 1 & -1.915 & 0.132 & 0.055 & -0.097 & 0.172 & 0.870 \\
& 1 & -1.915 & -0.132 & 0.055 & 0.097 & 0.172 & 0.870 \\
& 3 & -0.795 & -0.000 & -0.202 & -0.000 & 0.202 & 1.000 \\
& 3 & -0.795 & 0.000 & 0.202 & 0.000 & 0.202 & 1.000 \\
\hline
Inter-layer & 6 & 17.500 & -0.000 & -0.438 & 0.000 & 0.438 & 0.333 \\
interactions& 7 & 44.387 & 0.000 & 0.384 & -0.000 & 0.384 & 0.334 \\
& 5 & 12.908 & 0.166 & -0.016 & -0.044 & 0.172 & 0.508 \\
& 5 & 12.908 & -0.166 & -0.016 & 0.044 & 0.172 & 0.508 \\
& 8 & 2.947 & 0.023 & 0.158 & -0.053 & 0.168 & 0.697 \\
& 8 & -2.512 & -0.033 & -0.118 & 0.148 & 0.192 & 0.697 \\
& 8 & 2.947 & -0.023 & 0.158 & 0.053 & 0.168 & 0.697 \\
& 8 & -2.512 & 0.033 & -0.118 & -0.148 & 0.192 & 0.697 \\
& 6 & 2.375 & -0.000 & -0.006 & -0.000 & 0.006 & 0.782 \\
& 6 & 0.874 & -0.043 & 0.010 & 0.060 & 0.074 & 0.782 \\
& 6 & 0.874 & 0.043 & 0.010 & -0.060 & 0.074 & 0.782 \\
& 7 & 0.930 & 0.203 & 0.002 & -0.015 & 0.203 & 0.782 \\
& 7 & 0.930 & -0.203 & 0.002 & 0.015 & 0.203 & 0.782 \\
& 5 & -1.394 & -0.030 & -0.021 & 0.102 & 0.109 & 0.871 \\
& 5 & -1.394 & 0.030 & -0.021 & -0.102 & 0.109 & 0.871 \\
& 7 & 2.023 & 0.000 & -0.009 & -0.000 & 0.009 & 1.054 \\
\hline
\end{tabular}
\caption{\textbf{Heisenberg exchange interaction and DMI parameters for Fe 3 atom.} Both intra- and inter-layer interactions are taken into account. The distance between the neighbors is given in the units of the experimental lattice constant of InSb, 6.479 \AA.}
\label{Fe3}
\end{table}

\begin{table}[ht]
\centering
\begin{tabular}{|l|c|c|c|c|c|c|c|}
\hline
\multicolumn{8}{|c|}{Interactions for atom: Fe 4} \\
\hline
& Neighbor atom (Fe) & $J_{ij}$ (meV)& $D_x$ (meV)& $D_y$ (meV)& $D_z$ (meV)& $|\textbf{D}|$ (meV)& $|\textbf{R}|$ (a) \\
\hline
Intra-layer & 1 & 9.975 & -0.471 & 0.313 & 0.229 & 0.610 & 0.439 \\
interactions& 1 & 9.975 & 0.471 & 0.313 & -0.229 & 0.610 & 0.439 \\
& 2 & -0.545 & 0.000 & 0.461 & 0.000 & 0.461 & 0.508 \\
& 2 & -4.880 & -0.000 & -0.746 & 0.000 & 0.746 & 0.508 \\
& 3 & 2.537 & -0.214 & -0.057 & -0.028 & 0.223 & 0.612 \\
& 3 & 2.537 & 0.214 & -0.057 & 0.028 & 0.223 & 0.612 \\
& 3 & 4.395 & -0.207 & -0.035 & 0.259 & 0.334 & 0.612 \\
& 3 & 4.395 & 0.207 & -0.035 & -0.259 & 0.334 & 0.612 \\
& 4 & 1.519 & 0.113 & -0.000 & -0.079 & 0.138 & 0.707 \\
& 4 & 1.519 & -0.113 & 0.000 & 0.079 & 0.138 & 0.707 \\
& 1 & 1.168 & 0.070 & -0.215 & -0.058 & 0.233 & 0.832 \\
& 1 & 1.168 & -0.070 & -0.215 & 0.058 & 0.233 & 0.832 \\
& 2 & -2.272 & 0.039 & -0.074 & 0.058 & 0.102 & 0.871 \\
& 2 & -2.272 & -0.039 & -0.074 & -0.058 & 0.102 & 0.871 \\
& 2 & 0.140 & 0.070 & -0.117 & -0.081 & 0.158 & 0.871 \\
& 2 & 0.140 & -0.070 & -0.117 & 0.081 & 0.158 & 0.871 \\
& 4 & 1.745 & -0.000 & 0.026 & -0.000 & 0.026 & 1.000 \\
& 4 & 1.745 & 0.000 & -0.026 & 0.000 & 0.026 & 1.000 \\
\hline
Inter-layer & 8 & 38.832 & 0.000 & -0.805 & -0.000 & 0.805 & 0.323 \\
interactions& 5 & 17.941 & 0.000 & 0.506 & 0.000 & 0.506 & 0.358 \\
& 6 & 18.776 & -0.033 & 0.284 & 0.481 & 0.560 & 0.411 \\
& 6 & 18.776 & 0.033 & 0.284 & -0.481 & 0.560 & 0.411 \\
& 7 & 3.353 & -0.490 & -0.003 & 0.291 & 0.570 & 0.693 \\
& 7 & -2.174 & 0.124 & 0.077 & 0.289 & 0.324 & 0.693 \\
& 7 & -2.174 & -0.124 & 0.077 & -0.289 & 0.324 & 0.693 \\
& 7 & 3.353 & 0.490 & -0.003 & -0.291 & 0.570 & 0.693 \\
& 8 & -0.291 & 0.147 & -0.019 & -0.011 & 0.149 & 0.778 \\
& 8 & -0.291 & -0.147 & -0.019 & 0.011 & 0.149 & 0.778 \\
& 5 & -5.752 & -0.000 & 0.145 & 0.000 & 0.145 & 0.793 \\
& 5 & -0.857 & -0.010 & -0.128 & -0.087 & 0.155 & 0.793 \\
& 5 & -0.857 & 0.010 & -0.128 & 0.087 & 0.155 & 0.793 \\
& 8 & -2.708 & -0.000 & 0.115 & -0.000 & 0.115 & 1.051 \\
& 6 & 0.249 & 0.044 & -0.131 & -0.101 & 0.171 & 1.081 \\
& 6 & 0.249 & -0.044 & -0.131 & 0.101 & 0.171 & 1.081 \\
\hline
\end{tabular}
\caption{\textbf{Heisenberg exchange interaction and DMI parameters for Fe 4 atom.} Both intra- and inter-layer interactions are taken into account. The distance between the neighbors is given in the units of the experimental lattice constant of InSb, 6.479 \AA.}
\label{Fe4}
\end{table}

\begin{table}[ht]
\centering
\begin{tabular}{|l|c|c|c|c|c|c|c|}
\hline
\multicolumn{8}{|c|}{Interactions for atom: Fe 5} \\
\hline
& Neighbor atom (Fe) & $J_{ij}$ (meV)& $D_x$ (meV)& $D_y$ (meV)& $D_z$ (meV)& $|\textbf{D}|$ (meV)& $|\textbf{R}|$ (a) \\
\hline
Intra-layer & 6 & 10.630 & 0.096 & -0.184 & -0.090 & 0.226 & 0.436 \\
interactions& 6 & 10.630 & -0.096 & -0.184 & 0.090 & 0.226 & 0.436 \\
& 7 & 10.772 & 0.791 & 0.470 & -0.617 & 1.108 & 0.438 \\
& 7 & 10.772 & -0.791 & 0.470 & 0.617 & 1.108 & 0.438 \\
& 8 & 28.757 & 0.000 & -0.163 & -0.000 & 0.163 & 0.504 \\
& 8 & -4.314 & -0.000 & 0.100 & 0.000 & 0.100 & 0.504 \\
& 5 & -0.834 & -0.038 & -0.000 & 0.208 & 0.211 & 0.707 \\
& 5 & -0.834 & 0.038 & 0.000 & -0.208 & 0.211 & 0.707 \\
& 6 & -0.085 & -0.122 & 0.185 & -0.094 & 0.241 & 0.830 \\
& 6 & -0.085 & 0.122 & 0.185 & 0.094 & 0.241 & 0.830 \\
& 7 & -0.702 & 0.018 & -0.014 & -0.141 & 0.143 & 0.832 \\
& 7 & -0.702 & -0.018 & -0.014 & 0.141 & 0.143 & 0.832 \\
& 8 & -0.820 & 0.029 & -0.043 & -0.106 & 0.118 & 0.869 \\
& 8 & -0.820 & -0.029 & -0.043 & 0.106 & 0.118 & 0.869 \\
& 8 & -1.249 & 0.075 & -0.078 & -0.062 & 0.125 & 0.869 \\
& 8 & -1.249 & -0.075 & -0.078 & 0.062 & 0.125 & 0.869 \\
& 5 & -3.234 & 0.000 & -0.056 & 0.000 & 0.056 & 1.000 \\
& 5 & -3.234 & -0.000 & 0.056 & -0.000 & 0.056 & 1.000 \\
\hline
Inter-layer & 2 & 16.590 & -0.000 & -0.131 & 0.000 & 0.131 & 0.300 \\
interactions& 4 & 17.941 & -0.000 & -0.506 & -0.000 & 0.506 & 0.358 \\
& 3 & 12.908 & -0.166 & 0.016 & 0.044 & 0.172 & 0.508 \\
& 3 & 12.908 & 0.166 & 0.016 & -0.044 & 0.172 & 0.508 \\
& 1 & -6.111 & 0.186 & 0.071 & 0.042 & 0.204 & 0.639 \\
& 1 & -4.370 & -0.124 & -0.119 & 0.163 & 0.237 & 0.639 \\
& 1 & -4.370 & 0.124 & -0.119 & -0.163 & 0.237 & 0.639 \\
& 1 & -6.111 & -0.186 & 0.071 & -0.042 & 0.204 & 0.639 \\
& 2 & 0.688 & -0.040 & 0.015 & 0.030 & 0.052 & 0.768 \\
& 2 & 0.688 & 0.040 & 0.015 & -0.030 & 0.052 & 0.768 \\
& 2 & -9.858 & -0.000 & 0.265 & 0.000 & 0.265 & 0.768 \\
& 4 & -0.857 & 0.010 & 0.128 & 0.087 & 0.155 & 0.793 \\
& 4 & -0.857 & -0.010 & 0.128 & -0.087 & 0.155 & 0.793 \\
& 4 & -5.752 & 0.000 & -0.145 & -0.000 & 0.145 & 0.793 \\
& 3 & -1.394 & 0.030 & 0.021 & -0.102 & 0.109 & 0.871 \\
& 3 & -1.394 & -0.030 & 0.021 & 0.102 & 0.109 & 0.871 \\
& 2 & -0.484 & -0.013 & -0.097 & -0.039 & 0.105 & 1.044 \\
& 2 & -0.484 & 0.013 & -0.097 & 0.039 & 0.105 & 1.044 \\
\hline
\end{tabular}
\caption{\textbf{Heisenberg exchange interaction and DMI parameters for Fe 5 atom.} Both intra- and inter-layer interactions are taken into account. The distance between the neighbors is given in the units of the experimental lattice constant of InSb, 6.479 \AA.}
\label{Fe5}
\end{table}

\begin{table}[ht]
\centering
\begin{tabular}{|l|c|c|c|c|c|c|c|}
\hline
\multicolumn{8}{|c|}{Interactions for atom: Fe 6} \\
\hline
& Neighbor atom (Fe) & $J_{ij}$ (meV)& $D_x$ (meV)& $D_y$ (meV)& $D_z$ (meV)& $|\textbf{D}|$ (meV)& $|\textbf{R}|$ (a) \\
\hline
Intra-layer & 5 & 10.630 & 0.096 & 0.184 & -0.090 & 0.226 & 0.436 \\
interactions& 5 & 10.630 & -0.096 & 0.184 & 0.090 & 0.226 & 0.436 \\
& 8 & 11.922 & 0.099 & -0.340 & 0.030 & 0.355 & 0.448 \\
& 8 & 11.922 & -0.099 & -0.340 & -0.030 & 0.355 & 0.448 \\
& 7 & 18.670 & 0.000 & 0.163 & -0.000 & 0.163 & 0.513 \\
& 7 & -5.512 & 0.000 & -0.104 & 0.000 & 0.104 & 0.513 \\
& 6 & -4.925 & 0.193 & -0.000 & 0.196 & 0.275 & 0.707 \\
& 6 & -4.925 & -0.193 & 0.000 & -0.196 & 0.275 & 0.707 \\
& 5 & -0.085 & 0.122 & -0.185 & 0.094 & 0.241 & 0.830 \\
& 5 & -0.085 & -0.122 & -0.185 & -0.094 & 0.241 & 0.830 \\
& 8 & 0.513 & -0.039 & -0.152 & -0.327 & 0.363 & 0.837 \\
& 8 & 0.513 & 0.039 & -0.152 & 0.327 & 0.363 & 0.837 \\
& 7 & -3.173 & 0.026 & 0.092 & 0.213 & 0.234 & 0.873 \\
& 7 & -3.173 & -0.026 & 0.092 & -0.213 & 0.234 & 0.873 \\
& 7 & -0.553 & 0.009 & 0.028 & 0.032 & 0.043 & 0.873 \\
& 7 & -0.553 & -0.009 & 0.028 & -0.032 & 0.043 & 0.873 \\
& 6 & -0.560 & -0.000 & -0.126 & -0.000 & 0.126 & 1.000 \\
& 6 & -0.560 & 0.000 & 0.126 & 0.000 & 0.126 & 1.000 \\
\hline
Inter-layer & 1 & 9.022 & 0.000 & 0.323 & 0.000 & 0.323 & 0.284 \\
interactions& 3 & 17.500 & 0.000 & 0.438 & -0.000 & 0.438 & 0.333 \\
& 4 & 18.776 & 0.033 & -0.284 & -0.481 & 0.560 & 0.411 \\
& 4 & 18.776 & -0.033 & -0.284 & 0.481 & 0.560 & 0.411 \\
& 2 & -0.438 & -0.044 & 0.027 & 0.083 & 0.098 & 0.624 \\
& 2 & -0.438 & 0.044 & 0.027 & -0.083 & 0.098 & 0.624 \\
& 2 & 0.354 & 0.111 & 0.298 & -0.056 & 0.323 & 0.624 \\
& 2 & 0.354 & -0.111 & 0.298 & 0.056 & 0.323 & 0.624 \\
& 1 & 0.202 & -0.029 & -0.047 & -0.098 & 0.113 & 0.762 \\
& 1 & 0.202 & 0.029 & -0.047 & 0.098 & 0.113 & 0.762 \\
& 1 & -9.165 & 0.000 & 0.016 & -0.000 & 0.016 & 0.762 \\
& 3 & 0.874 & -0.043 & -0.010 & 0.060 & 0.074 & 0.782 \\
& 3 & 0.874 & 0.043 & -0.010 & -0.060 & 0.074 & 0.782 \\
& 3 & 2.375 & 0.000 & 0.006 & 0.000 & 0.006 & 0.782 \\
& 1 & -0.878 & -0.013 & -0.100 & -0.127 & 0.163 & 1.040 \\
& 1 & -0.878 & 0.013 & -0.100 & 0.127 & 0.163 & 1.040 \\
& 4 & 0.249 & 0.044 & 0.131 & -0.101 & 0.171 & 1.081 \\
& 4 & 0.249 & -0.044 & 0.131 & 0.101 & 0.171 & 1.081 \\
\hline
\end{tabular}
\caption{\textbf{Heisenberg exchange interaction and DMI parameters for Fe 6 atom.} Both intra- and inter-layer interactions are taken into account. The distance between the neighbors is given in the units of the experimental lattice constant of InSb, 6.479 \AA.}
\label{Fe6}
\end{table}

\begin{table}[ht]
\centering
\begin{tabular}{|l|c|c|c|c|c|c|c|}
\hline
\multicolumn{8}{|c|}{Interactions for atom: Fe 7} \\
\hline
& Neighbor atom (Fe) & $J_{ij}$ (meV)& $D_x$ (meV)& $D_y$ (meV)& $D_z$ (meV)& $|\textbf{D}|$ (meV)& $|\textbf{R}|$ (a) \\
\hline
Intra-layer & 5 & 10.772 & 0.791 & -0.470 & -0.617 & 1.108 & 0.438 \\
interactions& 5 & 10.772 & -0.791 & -0.470 & 0.617 & 1.108 & 0.438 \\
& 6 & 18.670 & -0.000 & -0.163 & 0.000 & 0.163 & 0.513 \\
& 6 & -5.512 & -0.000 & 0.104 & -0.000 & 0.104 & 0.513 \\
& 8 & -0.472 & 0.020 & 0.142 & -0.018 & 0.144 & 0.612 \\
& 8 & 0.059 & 0.512 & 0.536 & 0.308 & 0.802 & 0.612 \\
& 8 & -0.472 & -0.020 & 0.142 & 0.018 & 0.144 & 0.612 \\
& 8 & 0.059 & -0.512 & 0.536 & -0.308 & 0.802 & 0.612 \\
& 7 & 1.005 & -0.214 & -0.000 & 0.221 & 0.307 & 0.707 \\
& 7 & 1.005 & 0.214 & 0.000 & -0.221 & 0.307 & 0.707 \\
& 5 & -0.702 & -0.018 & 0.014 & 0.141 & 0.143 & 0.832 \\
& 5 & -0.702 & 0.018 & 0.014 & -0.141 & 0.143 & 0.832 \\
& 6 & -0.553 & 0.009 & -0.028 & 0.032 & 0.043 & 0.873 \\
& 6 & -0.553 & -0.009 & -0.028 & -0.032 & 0.043 & 0.873 \\
& 6 & -3.173 & 0.026 & -0.092 & 0.213 & 0.234 & 0.873 \\
& 6 & -3.173 & -0.026 & -0.092 & -0.213 & 0.234 & 0.873 \\
& 7 & 0.571 & 0.000 & 0.018 & -0.000 & 0.018 & 1.000 \\
& 7 & 0.571 & -0.000 & -0.018 & 0.000 & 0.018 & 1.000 \\
\hline
Inter-layer & 3 & 44.387 & -0.000 & -0.384 & 0.000 & 0.384 & 0.334 \\
interactions& 1 & 29.892 & 0.000 & -0.628 & -0.000 & 0.628 & 0.353 \\
& 2 & 10.437 & 0.248 & -0.037 & 0.037 & 0.253 & 0.424 \\
& 2 & 10.437 & -0.248 & -0.037 & -0.037 & 0.253 & 0.424 \\
& 4 & -2.174 & 0.124 & -0.077 & 0.289 & 0.324 & 0.693 \\
& 4 & -2.174 & -0.124 & -0.077 & -0.289 & 0.324 & 0.693 \\
& 4 & 3.353 & 0.490 & 0.003 & -0.291 & 0.570 & 0.693 \\
& 4 & 3.353 & -0.490 & 0.003 & 0.291 & 0.570 & 0.693 \\
& 3 & 0.930 & 0.203 & -0.002 & -0.015 & 0.203 & 0.782 \\
& 3 & 0.930 & -0.203 & -0.002 & 0.015 & 0.203 & 0.782 \\
& 1 & -4.700 & 0.000 & 0.033 & 0.000 & 0.033 & 0.790 \\
& 1 & 0.378 & -0.204 & -0.185 & -0.346 & 0.442 & 0.790 \\
& 1 & 0.378 & 0.204 & -0.185 & 0.346 & 0.442 & 0.790 \\
& 3 & 2.023 & -0.000 & 0.009 & 0.000 & 0.009 & 1.054 \\
& 2 & 0.486 & 0.019 & -0.086 & -0.113 & 0.144 & 1.086 \\
& 2 & 0.486 & -0.019 & -0.086 & 0.113 & 0.144 & 1.086 \\
\hline
\end{tabular}
\caption{\textbf{Heisenberg exchange interaction and DMI parameters for Fe 7 atom.} Both intra- and inter-layer interactions are taken into account. The distance between the neighbors is given in the units of the experimental lattice constant of InSb, 6.479 \AA.}
\label{Fe7}
\end{table}

\begin{table}[ht]
\centering
\begin{tabular}{|l|c|c|c|c|c|c|c|}
\hline
\multicolumn{8}{|c|}{Interactions for atom: Fe 8} \\
\hline
& Neighbor atom (Fe) & $J_{ij}$ (meV)& $D_x$ (meV)& $D_y$ (meV)& $D_z$ (meV)& $|\textbf{D}|$ (meV)& $|\textbf{R}|$ (a) \\
\hline
Intra-layer & 6 & 11.922 & -0.099 & 0.340 & -0.030 & 0.355 & 0.448 \\
interactions& 6 & 11.922 & 0.099 & 0.340 & 0.030 & 0.355 & 0.448 \\
& 5 & -4.314 & 0.000 & -0.100 & -0.000 & 0.100 & 0.504 \\
& 5 & 28.757 & -0.000 & 0.163 & 0.000 & 0.163 & 0.504 \\
& 7 & 0.059 & -0.512 & -0.536 & -0.308 & 0.802 & 0.612 \\
& 7 & 0.059 & 0.512 & -0.536 & 0.308 & 0.802 & 0.612 \\
& 7 & -0.472 & 0.020 & -0.142 & -0.018 & 0.144 & 0.612 \\
& 7 & -0.472 & -0.020 & -0.142 & 0.018 & 0.144 & 0.612 \\
& 8 & -4.057 & -0.451 & -0.000 & 0.013 & 0.452 & 0.707 \\
& 8 & -4.057 & 0.451 & 0.000 & -0.013 & 0.452 & 0.707 \\
& 6 & 0.513 & -0.039 & 0.152 & -0.327 & 0.363 & 0.837 \\
& 6 & 0.513 & 0.039 & 0.152 & 0.327 & 0.363 & 0.837 \\
& 5 & -1.249 & -0.075 & 0.078 & 0.062 & 0.125 & 0.869 \\
& 5 & -1.249 & 0.075 & 0.078 & -0.062 & 0.125 & 0.869 \\
& 5 & -0.820 & -0.029 & 0.043 & 0.106 & 0.118 & 0.869 \\
& 5 & -0.820 & 0.029 & 0.043 & -0.106 & 0.118 & 0.869 \\
& 8 & -1.024 & 0.000 & 0.019 & 0.000 & 0.019 & 1.000 \\
& 8 & -1.024 & -0.000 & -0.019 & -0.000 & 0.019 & 1.000 \\
\hline
Inter-layer & 4 & 38.832 & -0.000 & 0.805 & 0.000 & 0.805 & 0.323 \\
interactions& 2 & 39.792 & 0.000 & -0.408 & -0.000 & 0.408 & 0.342 \\
& 1 & 8.547 & 0.067 & -0.349 & 0.345 & 0.496 & 0.433 \\
& 1 & 8.547 & -0.067 & -0.349 & -0.345 & 0.496 & 0.433 \\
& 3 & -2.512 & 0.033 & 0.118 & -0.148 & 0.192 & 0.697 \\
& 3 & -2.512 & -0.033 & 0.118 & 0.148 & 0.192 & 0.697 \\
& 3 & 2.947 & 0.023 & -0.158 & -0.053 & 0.168 & 0.697 \\
& 3 & 2.947 & -0.023 & -0.158 & 0.053 & 0.168 & 0.697 \\
& 4 & -0.291 & 0.147 & 0.019 & -0.011 & 0.149 & 0.778 \\
& 4 & -0.291 & -0.147 & 0.019 & 0.011 & 0.149 & 0.778 \\
& 2 & -6.573 & -0.000 & -0.109 & -0.000 & 0.109 & 0.785 \\
& 2 & -0.703 & -0.185 & 0.104 & -0.012 & 0.213 & 0.785 \\
& 2 & -0.703 & 0.185 & 0.104 & 0.012 & 0.213 & 0.785 \\
& 4 & -2.708 & 0.000 & -0.115 & 0.000 & 0.115 & 1.051 \\
& 1 & 0.438 & -0.124 & -0.076 & 0.163 & 0.218 & 1.090 \\
& 1 & 0.438 & 0.124 & -0.076 & -0.163 & 0.218 & 1.090 \\
\hline
\end{tabular}
\caption{\textbf{Heisenberg exchange interaction and DMI parameters for Fe 8 atom.} Both intra- and inter-layer interactions are taken into account. The distance between the neighbors is given in the units of the experimental lattice constant of InSb, 6.479 \AA.}
\label{Fe8}
\end{table}

\newpage

\noindent\textbf{Magnetic characterization of the heterostructure through atomistic spin-lattice simulations} 

\noindent The Hamiltonian describing the magnetic ground state of the system is provided in Eq.~(6), as described in the Methods section.
To isolate the role of exchange frustration, we have initially removed all other interaction parameters, including DMI, magnetocrystalline anisotropy, and the Zeeman energy terms from the full Hamiltonian within our Monte Carlo (MC) simulations.
This approach allowed us to investigate whether competing exchange interactions alone can drive the system to an SS state solution. 
Remarkably, this simulation resulted in the spontaneous formation of an SS state driven entirely by competing exchange interactions $i.e.$, the exchange frustration. 
Analysis of the simulation data reveals an exchange frustration-driven spiral with a period of approximately 2.6 nm, as shown in Fig.~\ref{SS_DMI_Effect}\textbf{a}. 
Heisenberg exchange-frustrated systems with an SS order invariably acquire a chiral behavior (a distinct rotational sense) when DMI interactions are introduced into the Hamiltonian. 
Crucially, the DMI interaction dictates the handedness of the chiral order and exerts a subtle influence on the spiral's period, as shown in Fig.~\ref{SS_DMI_Effect}\textbf{b}.
In this cycloidal-SS state, the introduction of out-of-plane magnetocrystalline anisotropy can further reduce the SS period. 
As shown in Fig.~\ref{SS_DMI_Effect}\textbf{b}, the SS period exhibits a notable decrease of approximately 0.3 nm compared to the case driven solely by exchange frustration.

\noindent\textbf{Spontaneous nucleation of both elongated skyrmions and antiskyrmions.} 

\noindent Building upon the findings (cluster of skyrmions and antiskyrmions) presented in Extended Data Fig.~\textbf{5}, we here investigate the stability of metastable states characterized by the coexistence of skyrmions and antiskyrmions. 
To nucleate topological magnetic spin textures, our MC simulations begin with a random magnetic configuration and subsequently undergo a systematic process of simulated annealing at a finite magnetite field~\cite{sim_ann}.
In this process, the simulation domain involves two magnetic layers, each consisting of an $80 \times 80 \times1$ array of spins, subjected to a perpendicular magnetic field of approximately 0.5 Tesla. The total number of spins within the domain is 12800.
As presented in Fig.~\ref{Sponte_Sk_ASk_rand}, we observe the nucleation and subsequent evolution of topological charges with a magnitude of unity.
This leads to a domain exclusively composed of elongated skyrmions and antiskyrmions.
Notably, at zero temperatures, as depicted in Fig.~\ref{Sponte_Sk_ASk_rand}\textbf{d}, skyrmions and antiskyrmions are distinctly identified by red and blue boxes, respectively.
It is crucial to note that the true ground state at this magnetic field is the cone-SS phase, as evident in the magnetic phase diagram shown in Fig.~\textbf{4c} of the main text. Within this cone-SS background, a metastable cluster of skyrmions and antiskyrmions coexists.
While the precise number of skyrmions and antiskyrmions may vary, our simulations have not revealed any other topological spin textures with a topological charge greater than unity.
To further investigate this coexisting behavior, we employ a systematic simulated annealing protocol within MC simulations under a finite magnetic field. Now, for these simulations, we initialize the system with a cycloidal-SS configuration.
Figure~\ref{Sponte_Sk_ASk_SS} illustrates the results of our simulations, demonstrating a metastable state characterized by the coexistence of topologically distinct spin textures{--}skyrmions and antiskyrmions{--}within a background of cone-SS.
These findings provide critical support for the congruence between our micromagnetic and atomistic models, further validating our predictions for this novel magnetic phase in a real-material system.

\begin{figure*}[htbp]
    \centering    
    \includegraphics[width=1\linewidth]{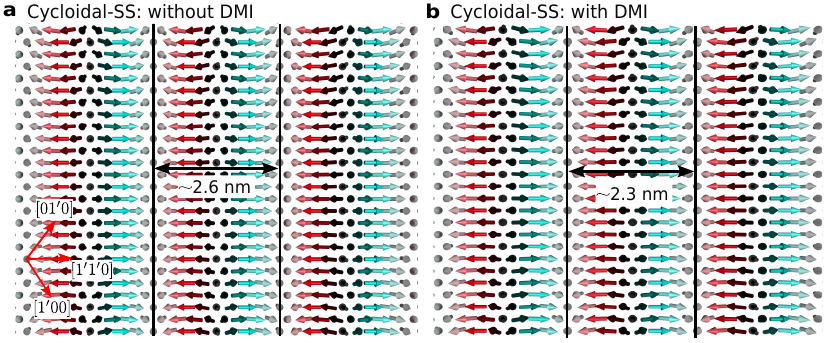}
    \caption{\textbf{Emergent spin spiral states: frustrated vs. chiral.} \textbf{a,} A spontaneous SS solution in the absence of DMI and magnetocrystalline anisotropy. \textbf{b,} A cycloidal-SS structure with DMI and magnetocrystalline anisotropy. Ultimately, the ground state in the absence of a magnetic field is a left-handed cycloidal-SS characterized by an atomic-scale period. To enhance visual clarity, only the top Fe layer is illustrated in these figures. }
    \label{SS_DMI_Effect}
\end{figure*}

\begin{figure*}[ht]
    \centering
    \includegraphics[width=0.86\textwidth]{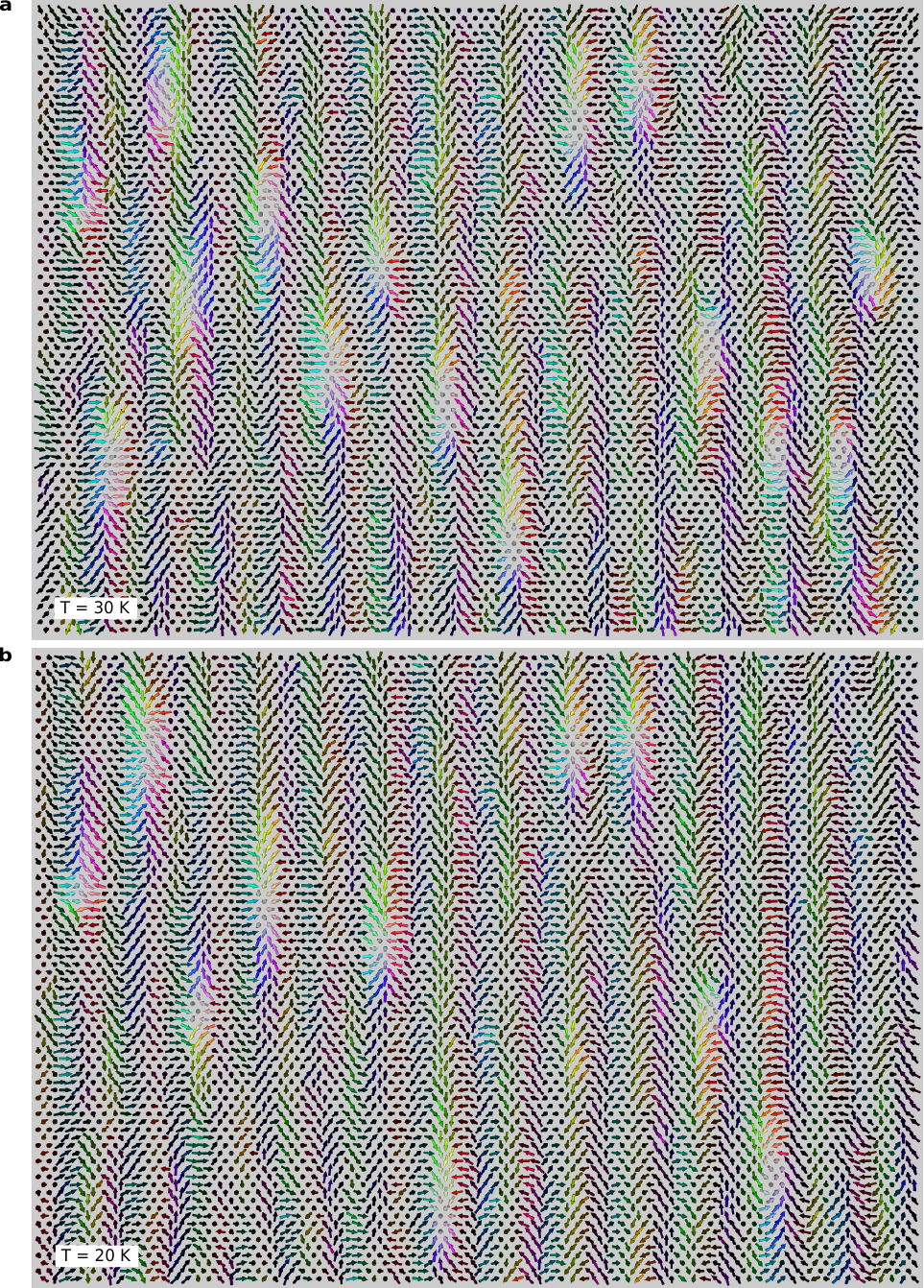}
\end{figure*}

\begin{figure*}[ht]
    \centering
    \includegraphics[width=0.83\textwidth]{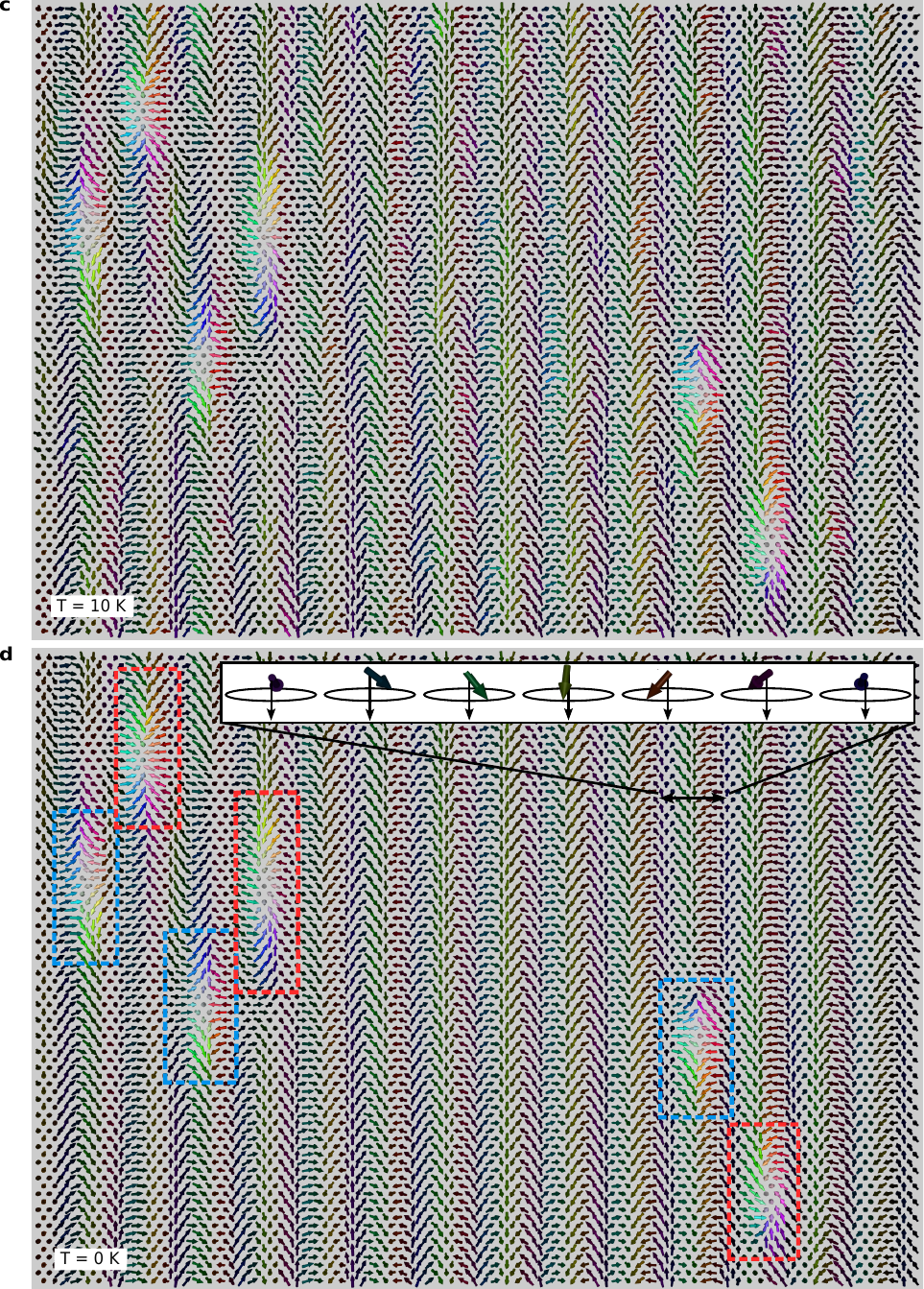}
    \caption{\textbf{Magnetic field induced spontaneous nucleation of both skyrmions and antiskyrmions in a finite domain.} At a finite temperature of about 50 K, simulations are initiated with a random distribution of spin orientations across the lattice sites. The temperature is systematically reduced in 0.5 K steps. \textbf{a-d}, the spin configurations obtained from our simulations at temperatures 30 K, 20 K, 10 K, and 0 K, respectively. In \textbf{d}, skyrmions are highlighted with red boxes, while antiskyrmions are marked with blue boxes. The inset provides a magnified view of the underlying cone-SS background modulation.}
    \label{Sponte_Sk_ASk_rand}
\end{figure*}

\begin{figure*}[ht]
    \centering
    \includegraphics[width=0.86\textwidth]{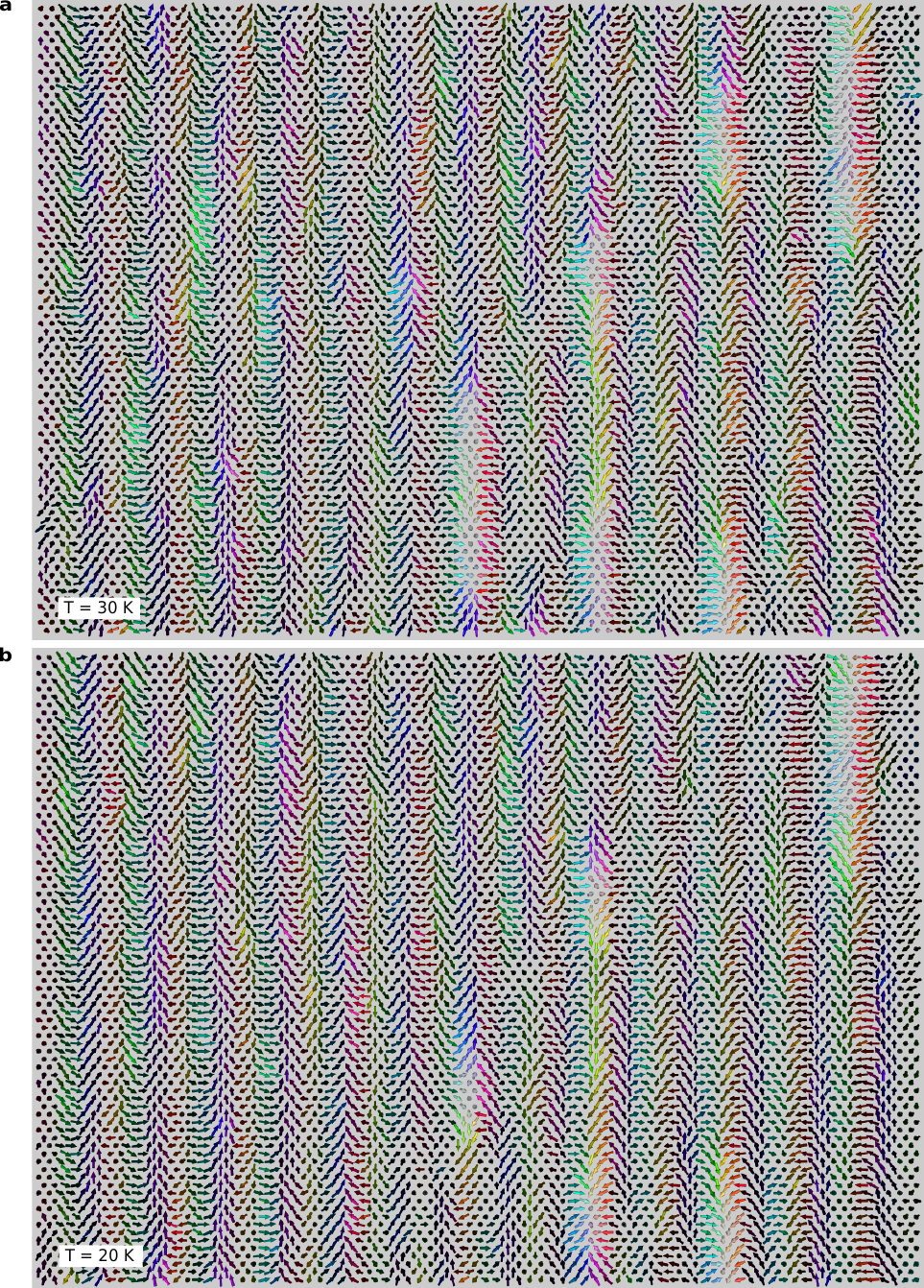}
\end{figure*}
\begin{figure*}[ht]
    \centering
    \includegraphics[width=0.83\textwidth]{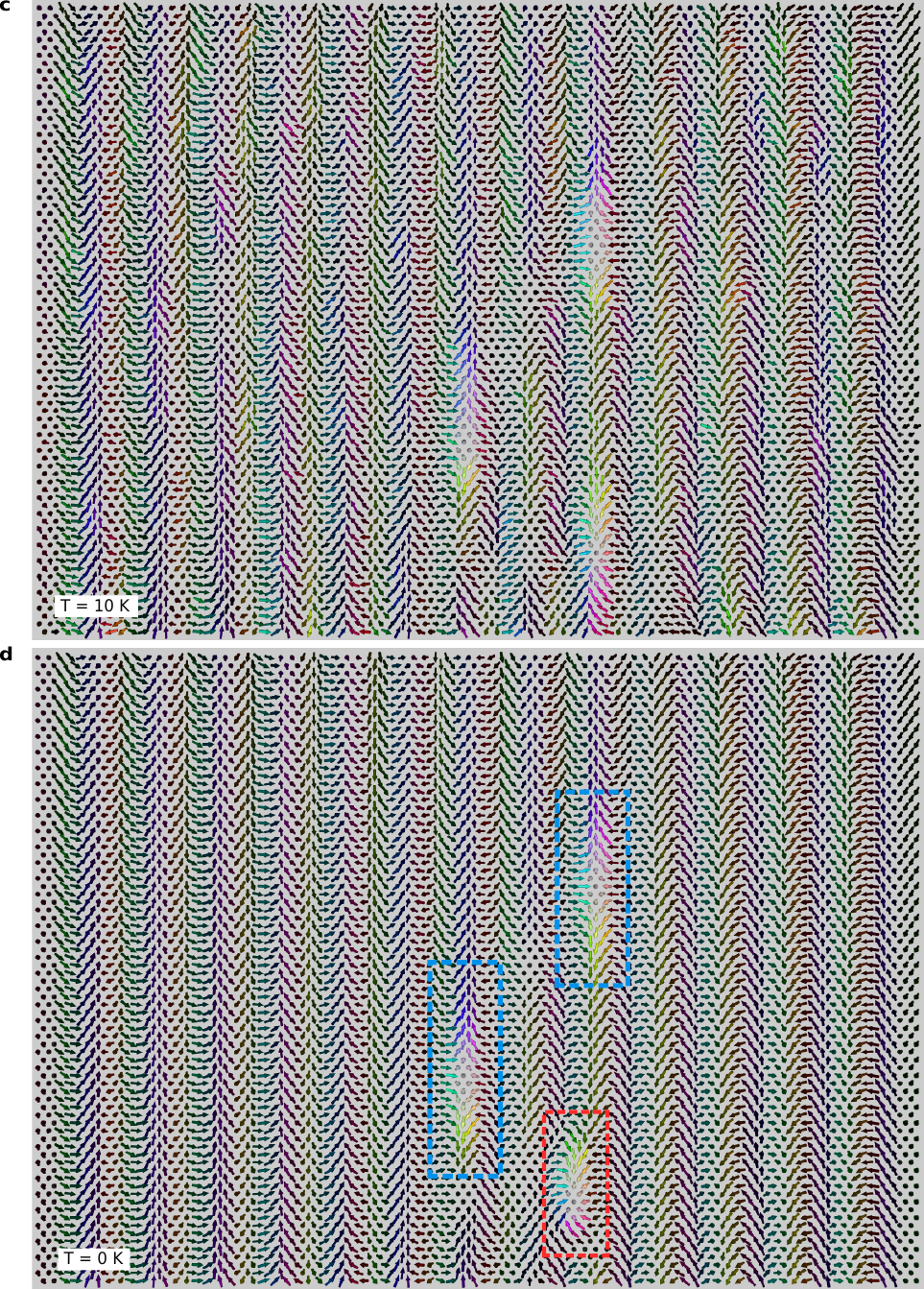}
    \caption{\textbf{Magnetic field induced spontaneous nucleation of both skyrmions and antiskyrmions in a finite domain initialized with cycloidal-SS state.} Following a similar procedure as in previous studies, but instead of employing a random spin configuration as the initial state, we initiate the simulations with the ground state cycloidal-SS configuration. \textbf{a-d}, the spin configurations obtained from our simulations at temperatures of 30 K, 20 K, 10 K, and 0 K, respectively. In \textbf{d}, skyrmions and antiskyrmions are highlighted with red and blue boxes, respectively.}
    \label{Sponte_Sk_ASk_SS}
\end{figure*}

\newpage

\begin{figure*}
\centering
\includegraphics[width=1.0\linewidth]{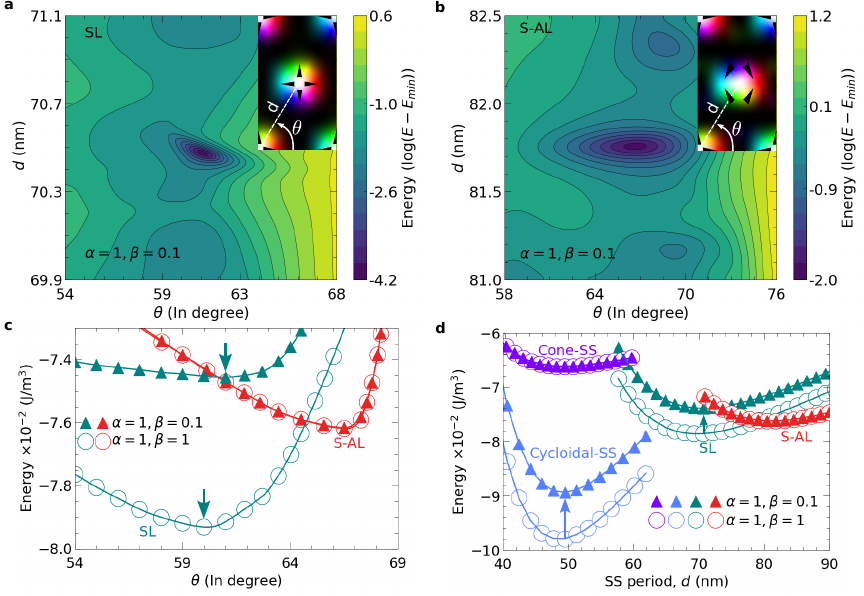}
\caption{\textbf{Energy optimization of noncollinear magnetic phases:} 
Contour color maps in panels \textbf{a} and \textbf{b} depict the energy density (per unit volume) of the SL and S-AL, respectively, as a function of the \textit{core-to-core} distance, $d$, and shape parameter, $\theta$. The inset illustrates the definition of these parameters within a rectangular unit cell.
All energy points are calculated relative to the minimum energy configuration, $E_\textrm{min}$, and presented on a logarithmic scale. The corresponding color variation is shown on the right.
Anisotropy in DMI magnitudes is considered, with $\beta$ fixed at 0.1.
\textbf{c,} Energy profile of the unit cell as a function of the shape parameter $\theta$, obtained through direct energy minimization.  
\textbf{d,} Energy density as a function of the \textit{core-to-core} distance $d$ within the SL and S-AL unit cells, and as a function of the period for the cone-SS and cycloidal-SS phases. The energy minima correspond to the equilibrium configurations.
Here, we have maintained a constant external magnetic field, $h=0.35$.
The optimized energy density for both S-AL and cone-SS phases remains invariant with respect to variations in the anisotropy parameter $\beta$.
However, upon introducing anisotropy in the DMI magnitude ($\beta$ = 0.1), we observe a significant increase in the energy density profiles of both the cycloidal-SS and SL phases compared to the isotropic case ($\beta$ = 1).
Despite the presence of DMI anisotropy, it is noteworthy that both optimized lattice states remain metastable regardless of the applied field. As $\beta=0.1 (< \beta_\textrm{c})$, S-AL becomes the energetically favorable phase compared to the SL. The optimized lattices for both phases are shown in the main text Figs.~\textbf{2a} and \textbf{b} for SL and S-AL, respectively.}
\label{fig:Extended1}
\end{figure*}

\begin{figure*}
	\centering
	\includegraphics[width=1\linewidth]{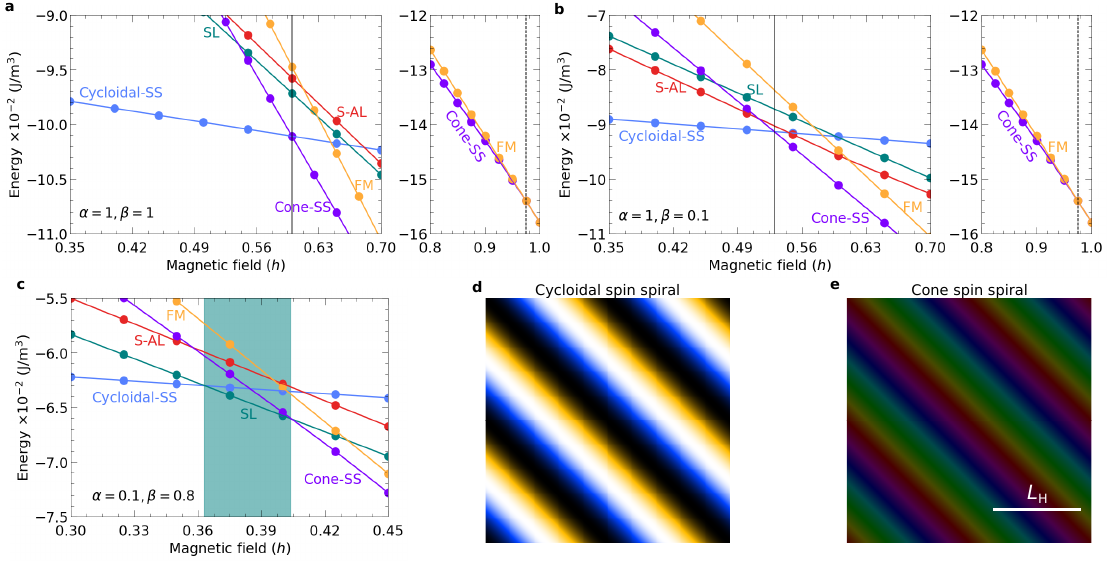}
	\caption{\textbf{Energy density vs. magnetic field of competing non-collinear magnetic phases:} 
This figure further extends our analysis by exploring the critical behavior of the anisotropy parameter $\beta$. This significantly influences the energy lines and the stability of various magnetic phases.  
\textbf{a,} Pure isotropic case. The energy densities shown in this figure are calculated using a simplified model of a frustrated chiral magnet, which omits the influence of magnetocrystalline anisotropy. 
The cycloidal-SS state remains the ground state for applied fields up to approximately 0.6. Beyond this critical field, the cone-SS emerges as the energetically favorable state.  
As depicted in the rightmost figure, a second-order phase transition occurs from the cone-SS to the saturated FM phase at high magnetic fields, indicated by the vertical line.
While both the SL and S-AL phases are metastable, the SL energy is always lower than that of S-AL. 
\textbf{b,} A case with anisotropy in DMI below the critical value, $\beta_\textrm{c}\sim 0.55$ (see Fig.~\ref{fig:Extended3}\textbf{c}). 
While the overall phase transition sequence remains consistent with the isotropic case, the introduction of anisotropy ($\beta=0.1$) leads to a subtle shift in energy within the lattice phases, favoring the S-AL over the SL.  
\textbf{c,} A scenario with both anisotropic interactions, DMI and exchange, $i.e.$, $\beta=0.8$ ($>\beta_\textrm{c}$) and $\alpha=0.1$.
In contrast to Fig.~\textbf{2e}, first-order phase transitions from the cycloidal-SS to the SL and subsequently from the SL to the cone-SS are observed at critical fields of approximately $h \approx$ 0.364 and 0.405, respectively.
Within the 2D domain, \textbf{d,} the cycloidal-SS and \textbf{e,} the cone-SS phase are determined through direct energy minimization of model~(\textbf{2}). Also, Figs.~\textbf{1f} and \textbf{g} illustrate the characteristic spin orientations of the cycloidal-SS and cone-SS phases, respectively.
}
	\label{fig:Extended2}
\end{figure*}

\begin{figure*}
\centering
\includegraphics[width=1\linewidth]{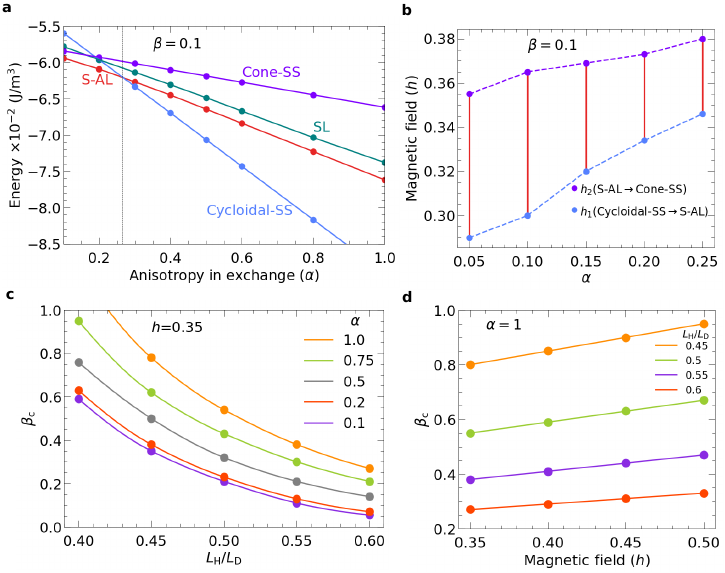}
\caption{\textbf{The role of $\alpha$ and $L_\textrm{H}/L_\textrm{D}$ in stabilizing S-AL phase:} \textbf{a,} Energy comparison of competing noncollinear states (cycloidal-SS, cone-SS, SL, and S-AL) as a function of the exchange anisotropy parameter $\alpha$, with $\beta$ fixed at 0.1. The vertical line marks the critical value of $\alpha$, below which the S-AL phase becomes energetically favorable magnetic state. 
For this simulation, a constant external magnetic field of $h = 0.35$ is applied, while keeping the ratio of $L_\textrm{H}/L_\textrm{D}$ fixed at 0.5.
\textbf{b,} Stability window of the ground-state S-AL phase as a function of the anisotropy parameter $\alpha$, below its critical value of 0.26. 
The magnetic fields $h_1$ and $h_2$ demarcate the phase boundaries between the cycloidal-SS, S-AL, and cone-SS phases. A notable correlation exists between $\alpha$ and the stability field window of the S-AL phase, with decreasing $\alpha$ resulting in an expanded window.
\textbf{c,} Dependence of $\beta_\textrm{c}$ for S-AL stability on various values of $L_\textrm{H}/L_\textrm{D}$ and $\alpha$. For values of $\beta$ below the critical value $\beta_\textrm{c}$, the S-AL can be emerged as the energetically favored ground state within a specific range of external magnetic fields. This range is bounded by two critical field values that define the transitions between the cycloidal-SS, S-AL, and cone-SS phases. Note, with $L_\mathrm{H}/L_\mathrm{D}=0.56$, $\beta_\textrm{c}$ is approximately 0.09. Therefore, the S-AL phase is absent as the ground state in Fig.~\textbf{3b} for all values of $h$.
\textbf{d,} Dependence of the critical parameter $\beta_\textrm{c}$, crucial for S-AL stability, on variations in $h$ for different $L_\textrm{H}/L_\textrm{D}$ ratios. Here, with $\alpha$ held constant at unity, we observed a linear relationship between $\beta_\textrm{c}$ and $h$.
}
	\label{fig:Extended3}
\end{figure*}

\begin{figure*}
	\centering
	\includegraphics[width=1\linewidth]{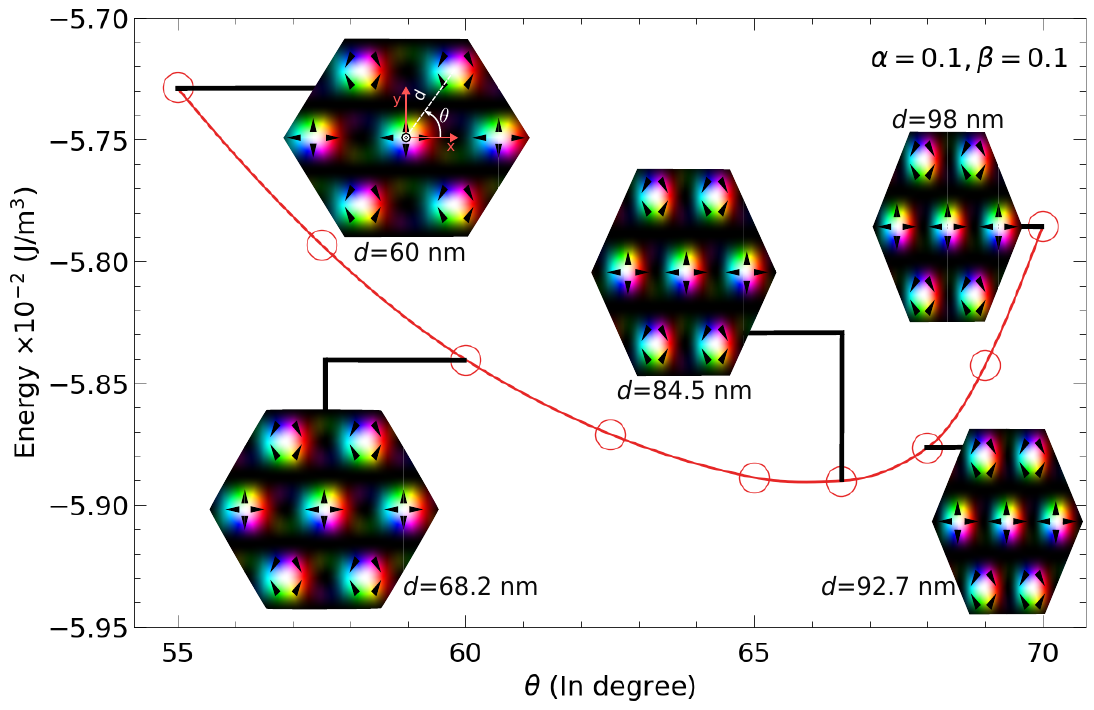}
	\caption{\textbf{Minimum energy S-AL with elongated skyrmion and antiskyrmion configurations:} Minimum energy density of the S-AL configuration determined by optimizing the \textit{core-to-core} distance ($d$) within a rectangular unit cell for each value of the fixed angle (shape parameter) $\theta$. For the anisotropic system with $\alpha=\beta=0.1$, the distance $d$ corresponds to the snapshot exhibiting minimum energy density. The equilibrium S-AL phase with the lowest energy is found at $\theta=66.5^\circ$, exhibiting elongated skyrmions and antiskyrmions as shown in Fig.~\textbf{2c}. 
    Here, our results unveil a significant correlation between the lattice shape and the skyrmion's elongation within the S-AL phase. Notably, the direction of elongation undergoes a marked change below $\theta=60^\circ$. For example, comparing S-AL configurations above and below this angle demonstrates a clear orthogonality in their elongation directions. This intriguing observation emphasizes the profound influence of DMI anisotropy on dictating the elongation direction. This also happens when we set $\alpha=1$. 
   }
	\label{fig:Extended4}
\end{figure*}

\begin{figure*}
	\centering
	\includegraphics[width=1\linewidth]{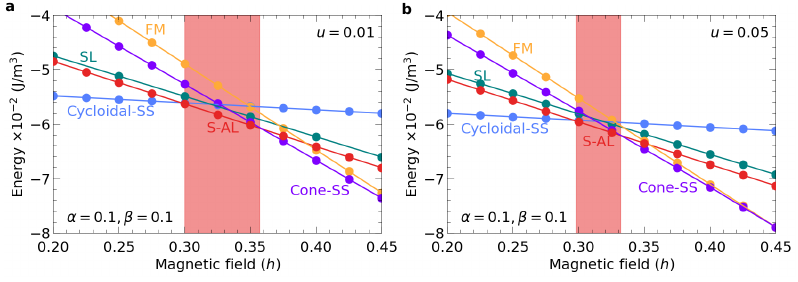}
	\caption{\textbf{Role of magnetocrystalline anisotropy parameter $u$:} By incorporating magnetocrystalline anisotropy as an additional energy term into our model~(\textbf{2}) (see Methods), we demonstrate here a more comprehensive understanding of 2D chiral magnets.
    \textbf{a} and \textbf{b} illustrate the impact of easy-axis anisotropy ($u$ = 0.01 and 0.05, respectively) on the stability of the S-AL phase.
    This extends the analysis presented in Fig.~\textbf{2g} to the case of non-zero easy-axis anisotropy ($u \ne 0$).
     Importantly, easy-axis anisotropy favors the cone-SS and saturated FM phases, leading to a second-order phase transition at lower magnetic fields with increasing $u$. 
     As a result, the stability range of the S-AL phase in terms of the external field decreases with increasing $u$.
}
	\label{fig:Extended5}
\end{figure*}

\begin{figure*}
    \centering
    \includegraphics[width=1.0\linewidth]{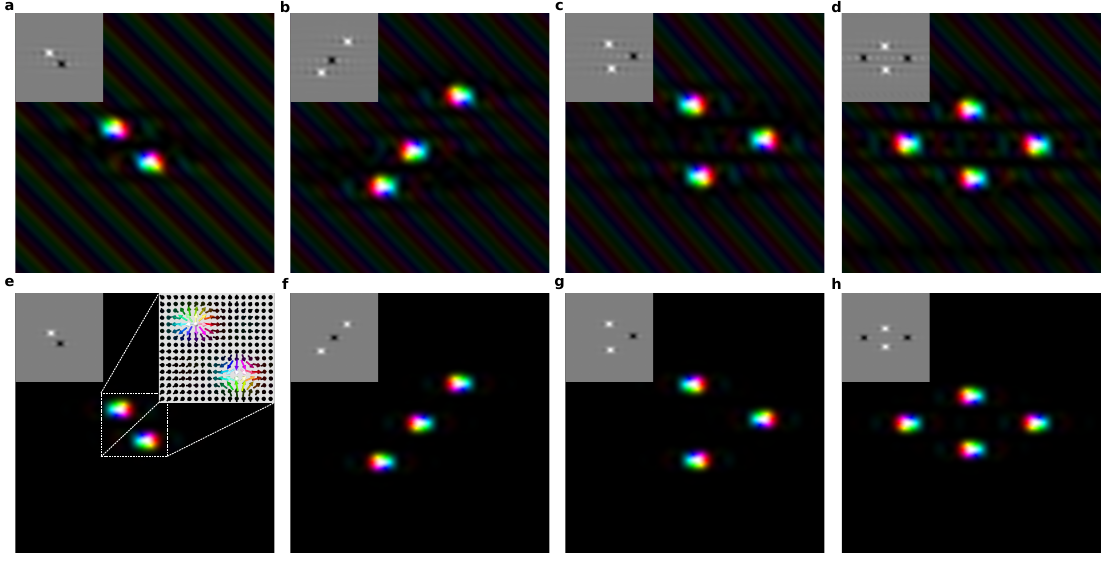}
\caption{\textbf{Stable skyrmion-antiskyrmion clusters in anisotropic chiral magnet:} 
Micromagnetic simulations within our frustrated chiral magnet with $\alpha=\beta=0.1$ reveal the formation of stable, multi-skyrmion clusters within a 2D domain. 
These clusters include: \textbf{a,} a skyrmion-antiskyrmion pair, \textbf{b,} linear and \textbf{c,} triangular arrangements of two skyrmions and one antiskyrmion, and \textbf{d,} a cluster of two skyrmions and two antiskyrmions.
Starting from initial configurations topologically equivalent to the desired final skyrmion cluster configurations, we have performed direct energy minimization of our model under an external magnetic field, $h=0.5$.
This minimization process is continued until a stable, minimum-energy configuration is reached.
All clusters appear to be embedded within the domain, which exhibits a cone-SS modulation in the background magnetization.
Upon increasing the magnetic field to $h=0.7$, the cone-SS background magnetization is suppressed, resulting in a homogeneous magnetization with skyrmion clusters unaltered, as shown in panels \textbf{e-h}.
The insets in each figure depict the topological charge density distribution within the domain. Concentrated black (white) dots against the gray background signify the presence of antiskyrmions (skyrmions) with topological charge $Q=1 (-1)$, respectively.
Upon closer examination in each case, the enlarged vector field (spin) representation reveals notable deviations in the shapes of individual skyrmions and antiskyrmions from their standard form. For example, see the inset of \textbf{e} for the skyrmion-antiskyrmion pair.
}
\label{fig:Extended6}
\end{figure*}
\clearpage
\end{onecolumngrid}
\end{document}